\documentclass[a4paper,fleqn,usenatbib,useAMS]{mnras}

\usepackage{mathptmx}

\usepackage[T1]{fontenc}
\usepackage{ae,aecompl}

\usepackage{etoolbox}
\makeatletter
\patchcmd\@combinedblfloats{\box\@outputbox}{\unvbox\@outputbox}{}{\errmessage{\noexpand patch failed}}
\makeatother
 
\usepackage{graphicx}	
\usepackage{amsmath}	
\usepackage{amssymb}	
\usepackage{multicol}   

\newcommand{\pcmm}{\,cm$^{-3}$}	
\newcommand{\kms}{\,km\,s$^{-1}$} 
\newcommand{\water}{H$_{2}$O} 
\newcommand{\form}{H$_{2}$CO} 
\newcommand{\meth}{CH$_{3}$OH} 
\newcommand{\ngc}{NGC~6334} 
\newcommand{\ngcf}{NGC~6334F} 
\newcommand{\ngci}{NGC~6334I} 
\newcommand{\UCHII}{UCH\,{\sc ii}} %

\makeatletter\ifdefined\Hy@StartlinkName
\def\addOneNestingLevelStartLink{%
  \gdef\Hy@StartlinkName##1##2{%
    \sbox0{\Hy@StartlinkNameOrig{##1}{##2}}\usebox0
    \global\let\Hy@StartlinkName\Hy@StartlinkNameOrig%
  }%
}
\def\addOneNestingLevelEndLink{%
  \gdef\pdfendlink{%
    \sbox0{\pdfendlinkOrig}\usebox0%
    \global\let\pdfendlink\pdfendlinkOrig%
  }%
}
\let\Hy@StartlinkNameOrig\Hy@StartlinkName
\let\pdfendlinkOrig\pdfendlink
\else
\let\addOneNestingLevelStartLink\relax
\let\addOneNestingLevelEndLink\relax
\fi

\title[A Masing Event in \ngci]{A Masing Event in \ngci: Contemporaneous Flaring of Hydroxyl, Methanol and Water Masers}

\author[G. C. MacLeod et al.]{
G. C. MacLeod$^{1}$\thanks{E-mail: gord@hartrao.ac.za}, D. P. Smits$^{2}$,
S. Goedhart$^{3}$, T. R. Hunter$^{4}$, C. L. Brogan$^{4}$, 
\newauthor{
J. O. Chibueze$^{3, 5, 6}$, S. P. van den Heever$^{1}$, C. J. Thesner$^{5}$, P. J. Banda$^{7}$, and}
\newauthor{
J. D. Paulsen$^{8}$ }\\
$^{1}$Hartebeesthoek Radio Astronomy Observatory, PO Box 443, Krugersdorp, 1741, 
South Africa \\
$^{2}$Dept of Mathematical Sciences, UNISA, PO Box 392, UNISA, 0003, South Africa \\
$^{3}$SKA SA, The Park, Park Road, Pinelands, South Africa \\
$^{4}$NRAO, 520 Edgemont Rd, Charlottesville, VA, 22903, USA \\
$^{5}$Space Research Unit, Physics Department, North West University, Potchefstroom, South Africa \\
$^{6}$Department of Physics and Astronomy, University of Nigeria, Carver Building, 1 University Road, Nsukka, Nigeria \\
$^{7}$School of Physical Sciences, University of Nairobi, Nairobi, Kenya \\
$^{8}$Trinity House High School, Little Falls, South Africa 
}

\date{Accepted 2018 April 4. Received 2018 March 22; in original form 2017 October 18}

\pubyear{2018}

\begin{document}
\label{firstpage}
\pagerange{\pageref{firstpage}--\pageref{lastpage}}
\maketitle

\begin{abstract}
As a product of the maser monitoring program with the 26\,m telescope of the Hartebeesthoek Radio Astronomy Observatory (HartRAO), we present an unprecedented, contemporaneous flaring event of 10 maser transitions in hydroxyl, methanol, and water that began in 2015 January in the massive star-forming region \ngci\ in the velocity range $-$10 to $-$2\kms. The 6.7\,GHz methanol and 22.2\,GHz water masers began flaring within 22~days of each other, while the 12.2\,GHz methanol and 1665\,MHz hydroxyl masers flared 80 and 113~days later respectively. The 1665\,MHz, 6.7\,GHz, and 22.2\,GHz masers have all remained in their flared state for nearly 3 years.  The brightest flaring components increased by factors of 66, 21, 26, and 20 in the 12.2 and 6.7\,GHz methanol, 1665\,MHz hydroxyl and 22.2\,GHz water maser transitions respectively; some weaker components increased by up to a factor of 145. We also report new maser emission  in the 1720, 6031, and 6035\,MHz OH lines and the 23.1\,GHz methanol line, along with the detection of only the fifth 4660\,MHz OH maser. We note the correlation of this event with the extraordinary (sub)millimeter continuum outburst from the massive protostellar system NGC~6334I-MM1 and discuss the implications of the observed time lags between different maser velocity components on the nature of the outburst.  Finally, we identify two earlier epoch maser flaring events likely associated with this object, which suggest a recurring accretive phenomenon that generates powerful radiative outbursts.
\end{abstract}

\section{Introduction}
Cosmic masers act as probes into heavily obscured astrophysical sources providing observers with information about the dynamics and physics of a region. Shortly after the discovery of masers in 1963 \citep{wwdl65}, temporal variability of the 1665 and 1667\, MHz groundstate OH maser lines in the star-forming complex \ngc\ was reported by \citet{wdw68}. Some velocity channels displayed variations by an order of magnitude between 1965 July and 1966 February. Evidently, their observations represented the first reported maser flare though they referred to it only as `a new phenomenon in microwave spectroscopy'.  In subsequent infrared continuum observations, \ngc\/ was identified as one of the most active regions of massive star-formation in the Galaxy \citep[see][and references therein]{hg83}.  More recent higher resolution, multi-wavelength studies have shown that it contains many young, deeply-embedded clusters and protoclusters \citep{Hunter14,Feigelson2009,PT08}, numerous massive dense cores and filaments \citep{Tige2017,Fukui17,Andre16,Russeil13}, and appears to be undergoing a `mini-starburst' event \citep{willis13}. 

\begin{table*}
\centering
\caption{Spectral transitions observed, monitoring start month, and receiver packages used at HartRAO.} 
\label{tab:transitions}
\begin{tabular}{cccccccccc}
\hline
Mol.  & Receiver & \multicolumn{2}{c}{Maser}   &Beam & Band &\multicolumn{2}{c}{Velocity}   &Sensitivity & Monitoring\\
        & &   Transition   &  Freq.     & Width &  Width$^{1}$  &Range         &Resolution      &3-$\sigma$ rms & Start\\
          & (cm) &  & (MHz)        &($\arcmin$)   &(MHz)     &(\kms)        &(\kms)          &(Jy) & Month\\  
\hline 
OH	& 18 & $^2\Pi_{J=3/2} ~F = 1 \rightarrow 2$ & 1612.231  & 29.6 & 0.25  & 22.5        & 0.045  & 0.4 & 2015 Sept     \\
	& 18 & $^2\Pi_{J=3/2} ~F = 1 \rightarrow 1$ & 1665.402  & 29.6 & 0.25$^{2}$ & 45.0  & 0.045  & $0.8 - 1.2$& 2011 Oct  \\
	& 18 & $^2\Pi_{J=3/2} ~F = 2 \rightarrow 2$ & 1667.359  & 29.6 & 0.25  & 22.5 & 0.044  & $0.4 - 0.8$& 2015 Sept \\
	& 18 & $^2\Pi_{J=3/2} ~F = 2 \rightarrow 1$ & 1720.530  & 29.6 & 0.25  & 22.5  & 0.043  & 0.4 & 2015 Sept      \\
	& 6   & $^2\Pi_{J=1/2} ~F = 0 \rightarrow 1$ & 4660.242  & 9.6 & 1.0   & 33.0  & 0.063 & $0.2 - 0.4$&  2015 Sept \\
	& 6   & $^2\Pi_{J=1/2} ~F = 1 \rightarrow 1$ & 4750.656  & 9.6 & 1.0 & 33.0 & 0.062 & 0.2 & 2015 Sept      \\
	& 6   & $^2\Pi_{J=1/2} ~F = 1 \rightarrow 0$ & 4765.562  & 9.6 & 1.0  & 33.0 &0.061  & $0.2 - 0.3$ &  2015 Sept\\
	& 5   & $^2\Pi_{J=5/2} ~F = 2 \rightarrow 3$ & 6016.746  & 7.5 & 1.0 & 24.5 & 0.049 & 0.2 & 2015 Sept      \\
	& 5   & $^2\Pi_{J=5/2} ~F = 2 \rightarrow 2$ & 6030.747$^{3}$  & 7.5 & 1.0   & 24.5 & 0.049  & 0.4&  2015 Sept \\
	& 5   & $^2\Pi_{J=5/2} ~F = 3 \rightarrow 3$ & 6035.092$^{3}$  & 7.5 & 1.0 & 24.5 & 0.049 & $0.3 - 0.4$& 2015 Aug \\
	& 5   & $^2\Pi_{J=5/2} ~F = 3 \rightarrow 2$ & 6049.084  & 7.5 & 1.0   & 24.5        & 0.048  &$0.2 - 0.3$& 2015 Sept \\
\form	 & 6  & $J = 1_{11} \rightarrow 1_{10}$  & 4829.660$^{4}$  & 9.6 & 1.0   & 33.0        &  0.060  & 0.4& 2016 Nov       \\
\meth  & 4.5 & $J = 5_{1} \rightarrow 6_{0} ~A^{+}$ & 6668.518  & 7.0 &0.64$^{5}$ & 14.5 &  0.112  & $1.0 - 1.5$ & 1999 Feb\\
       &     &   &  &   & 1.0 & 22.5 & 0.044 & $0.8 - 1.2$ & 2003 Mar\\
 	 & 2.5 & $J = 2_{0} \rightarrow 3_{-1} ~E$ & 12178.593$^{6}$ & 4.0 & 0.64$^{5}$  &  7.75 & 0.061 & $1.5 - 2.5$ & 2000 Jan\\
       &     &   &  &   & 2.0 & 24.5 & 0.048 & $0.6 - 0.9$& 2003 Mar\\
 	 & 1.3 & $J = 9_{2} \rightarrow 10_{1} ~A^{+}$ & 23121.024 & 2.2 & 8.0$^{2}$   & 107.9 & 0.101  & $1.1 - 2.3$& 2015 Aug\\
\water   & 1.3 & $J = 6_{16} \rightarrow 5_{23}$ &22235.120 & 2.2 & 8.0$^{2}$   & 107.9 & 0.105  & $2.3 - 2.9$& 2011 Apr \\
\hline
\multicolumn{10}{l}{$^{1}$A dual polarisation, 1024 channel each, spectrometer was employed after 2003 March 27.}\\
\multicolumn{10}{l}{$^{2}$The method of position switching was employed.}\\
\multicolumn{10}{l}{$^{3}$Monitoring was discontinued after 2016 June 01.}\\
\multicolumn{10}{l}{$^{4}$Only a single observation was made on 2016 November 12.}\\
\multicolumn{10}{l}{$^{5}$A single polarisation, 256 channel, spectrometer at LCP was employed until 2003 March 27.}\\
\multicolumn{10}{l}{$^{6}$The receiver was offline from 2016 January 01 to August 09.}\\
\end{tabular}
\end{table*}

Among the various centres of star formation in \ngc, one of the youngest contains a bright ultracompact H\,{\sc ii} (\UCHII) region \ngcf\/ \citep[G351.42+0.64,][]{rcm82,gm87,enm96}, which is associated with the far-infrared/millimeter source \ngci\ \citep{DeBuizer02,Gezari82,MFSW79,Cheung78,Emerson73}. The millimeter emission from this region has been resolved by interferometers into multiple continuum sources MM1$-$4 \citep{hbm06, bhccfi16}, two of which contain hot molecular cores accompanied by strong dust emission -- MM1 and MM2 \citep{bwtz07,zetal12,metal17,Bogelund18}. The combination of these four objects (hereafter referred to as \ngci, see Fig. \ref{fig:spots}) produce maser radiation from hydroxyl (OH), water (\water) at 22.2\,GHz \citep{mcbswb69}, and methanol (\meth) at 12.2\,GHz  \citep{bmmw87}, 6.7\,GHz \citep{m91} and various other higher frequencies \citep[see][]{valtts99,Haschick90,mb89,Haschick89,hb89}.  Many of these masers have been found to vary, as noted in several published works (see section~\ref{discussion}). Such variations have motivated more regular monitoring of multiple maser lines at the Hartebeesthoek Radio Astronomy Observatory \mbox{(HartRAO)}.

In this paper, we present the results of multi-year monitoring of 16 maser transitions (11 hydroxyl, 3 methanol, 1 formaldehyde, and  1 water) associated with \ngci\ since 1999. During 2015, 10 of these transitions started flaring, some of which have since dropped below our detection limits while others persist through 2017. We determine the onset of the flare in different velocity channels and when peaks occur, and discuss the possible physical causes of the flares and their association with the recent (sub)millimeter continuum outburst within the massive protostellar system NGC~6334I-MM1 detected with the Atacama Large Millimeter/submillimeter Array (ALMA) in July 2015 \citep{hetal17}. The methanol maser data from \citet{ggv04} have been re-assessed, leading to the identification of an earlier flare in 1999, possibly arising from the same object.  Finally, we make a prediction for the date of a future outburst and suggest further observations to promote the ongoing study of this very interesting massive star-forming region. 

\section{Observations}
The observations reported here were made using the 26\,m telescope of Hartebeesthoek Radio Astronomy Observatory (HartRAO)\footnote{See http://www.hartrao.ac.za/spectra/ for further information.}.  Information for each transition observed, and receiver used, is listed in Table~\ref{tab:transitions}. Typically, observations were made every 10 to 15\,d commencing during the start month listed in Table~\ref{tab:transitions}. However, the cadence of observations varied depending on the availability of the telescope, and the weather conditions. At times observations were done daily, but there are also observations separated by weeks. The coordinates that the telescope pointed to were R.A.~=~17$^{h}$~20$^{m}$~53$\fs$4 and Dec.~=~$-35\degr$~47$\arcmin$~01$\farcs5$ (J2000).

Prior to 2003 March 27, only left circularly polarised (LCP) feeds were installed on the telescope and spectra were obtained using a 256-channel spectrometer, as described in \citet{ggv04}. From 2003 March 27 onward, each receiver system consisted of left and right circularly polarised (RCP) feeds. Dual polarization spectra were obtained using a 1024-channel (per polarisation) spectrometer. The 2.5\,cm receiver operates at ambient temperature, the others are cryogenically cooled. Each polarisation  was calibrated independently relative to Hydra~A and 3C123 (and Jupiter in the case of the 1.3\,cm receiver), assuming the flux scale of \citet{Ott94}. Typical sensitivities achieved per observation are presented in Table~\ref{tab:transitions}. Observations made with the 1.3\,cm receiver were corrected for atmospheric absorption, the other observations were not (the effect is less than 3 per cent in these transitions).

\begin{figure*}   
	\includegraphics[width=\textwidth]{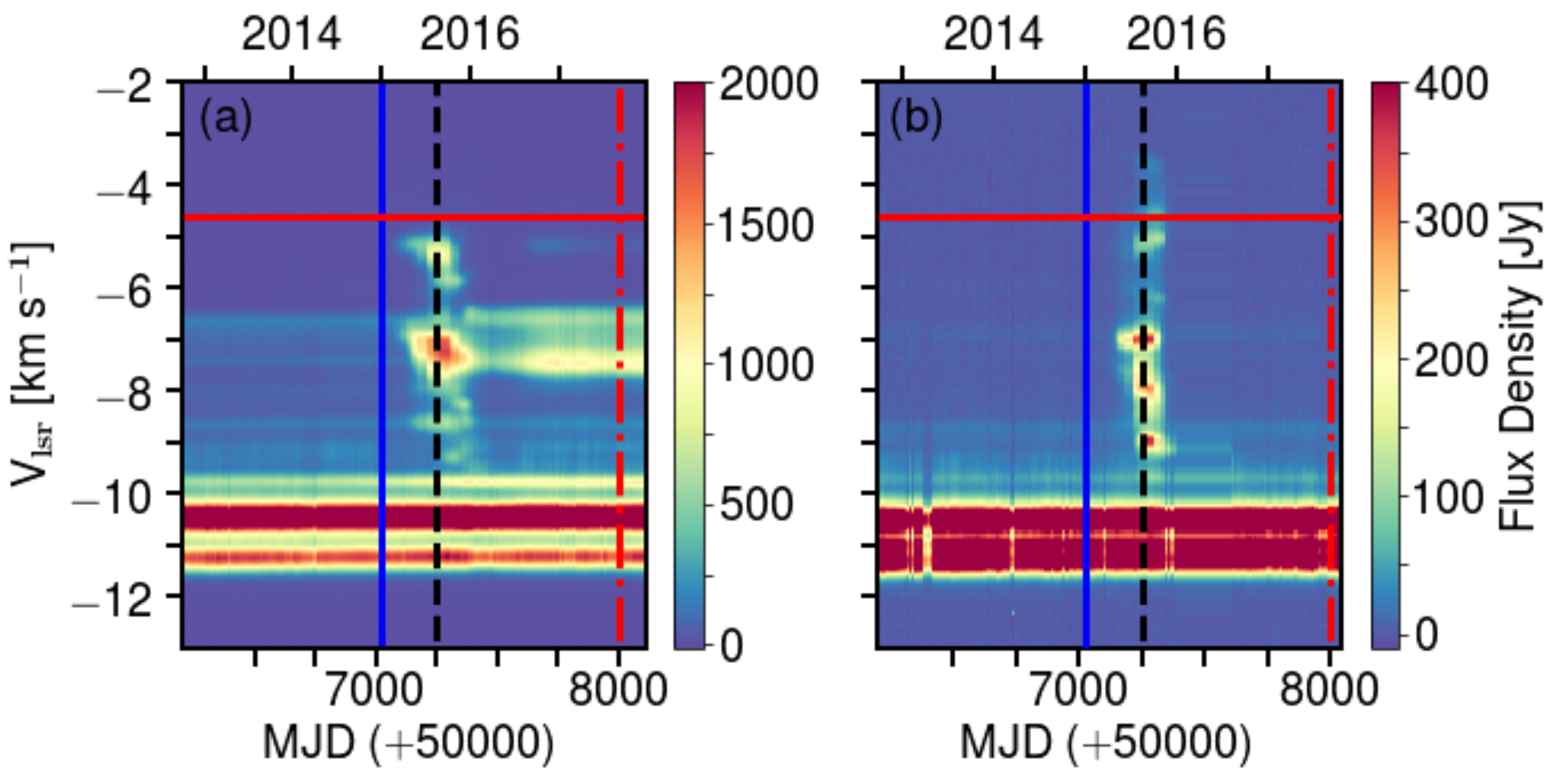}
	\caption{Dynamic spectra of a sub-set of the methanol maser observations for: (a) 6.7\,GHz, and (b) 12.2\,GHz. The solid (blue), dashed (black) and dot-dashed (red) vertical lines represent the estimated onset (2015 January 01/MJD 57023 = day 0), peak (2015 August 15/MJD 57249) and a recent epoch (2017 August 31/MJD 57997) of Kitty. The horizontal red line demarcates the last red shifted channel with 6.7\,GHz emission.}
 \label{fig:m67_ct}
\end{figure*}

Because of their large velocity extents, both \water\ and 1665\,MHz OH emissions were observed in position switching mode. Frequency switching was employed for observations at all other frequencies. Some contamination from an absorption feature at $+$6.4\kms\ was introduced into the 1667\,MHz maser emission spectra at $-$16\kms\ resulting from using frequency switching. The 23.1\,GHz \meth\ observations were initially done in position switching mode but were changed to frequency switching mode later.  Because of the poor sensitivity of the receiver system at 23.1\,GHz, several observations had to be averaged to obtain a sufficient signal-to-noise ratio. Pointing observations were performed for the 1665\,MHz OH transition on 2017 January 21; no pointing observations were observed for any other OH transition. Water and methanol observations included pointing observations. The 4829\,MHz formaldehyde transition was observed only once on 2016 November 12. Because there was no detection, no further observations were done. Monitoring of 6031 and 6035\,MHz OH masers was discontinued on 2016 June 01 because radio frequency interference became increasingly worse after 2016 February. No observations were taken between mid-2008 and early-2010 while the 26\,m antenna underwent repairs \citep{Gaylard10}. The 12.2\,GHz \meth\ receiver was offline for repairs between 2016 January 10 and August 09.

\section{Results}
We present the results of an unprecedented maser flaring event in \ngci\/ that started in 2015. Many of our times are presented using Modified Julian Date (MJD = Julian Date $-$ 240\,0000.5). Because \ngci\ is found in the Cat's Paw Nebula, we refer to these 10 flaring maser transitions (from three molecules) in 2015 collectively as `Kitty'. The transitions are 6.7, 12.2 and 23.1\,GHz of \meth, 1665, 1667, 1720, 4660, 6031, and 6035\,MHz of OH, and the 22.2\,GHz line of \water.  

\subsection{Methanol Masers}
The clearest evidence of the flaring event comes from the methanol data.  The archival regular monitoring data of the 6.7 and 12.2\,GHz methanol observations for \ngci\ were used to produce dynamic spectra, which are shown in Fig.~\ref{fig:m67_ct}. The flux density colour scales have been chosen to include the strongest components of the flare in each plot, and are consequently different for the 6.7 and 12.2\,GHz maps. Velocity channels with persistent maser emission are seen as horizontal trails across the image; hereafter, we refer to these as `contrails'. Contrails can vary in flux density ($F$) with time. Note that no interpolation is done between epochs of observation. 

The contrails in the 6.7\,GHz plot indicate that there are a number of velocity channels in which there are masers that are present prior to 2015.  Between $v = -10$ and $-11.5$\kms\/ there are some strong masers ($F(6.7) > 2\,000$\,Jy, where $F(6.7)$ is the flux density of the maser emission at 6.7\,GHz) which do vary but do not flare during this time period.  Between $v = -4.6$ (horizontal red line) and $-10$\kms\/ after the start of 2015 (MJD 57023, vertical blue line) a number of new velocity channels suddenly exhibit new emission, while some of the existing channels increase dramatically in brightness, all of which peak around 2015 August 15 (MJD 57249).  Eventually, the emission in most of these channels either disappears or returns to its quiescent level, but some channels undergo a re-brightening in 2016.

The 12.2\,GHz dynamic spectrum also has strong contrails between $v = -10$ and $-11.5$\kms, some weak contrails before 2015, and a group of velocity channels that brighten and reach a peak at a similar time to those at 6.7\,GHz. The 12.2\,GHz emission spans a larger velocity range than the 6.7\,GHz lines which is obvious by the presence of components above the red horizontal line. Note that although the scales are different, we checked the processed data and found a lack of 6.7\,GHz emission at these velocities. 

\begin{figure}   
	\includegraphics[clip,width=\columnwidth]{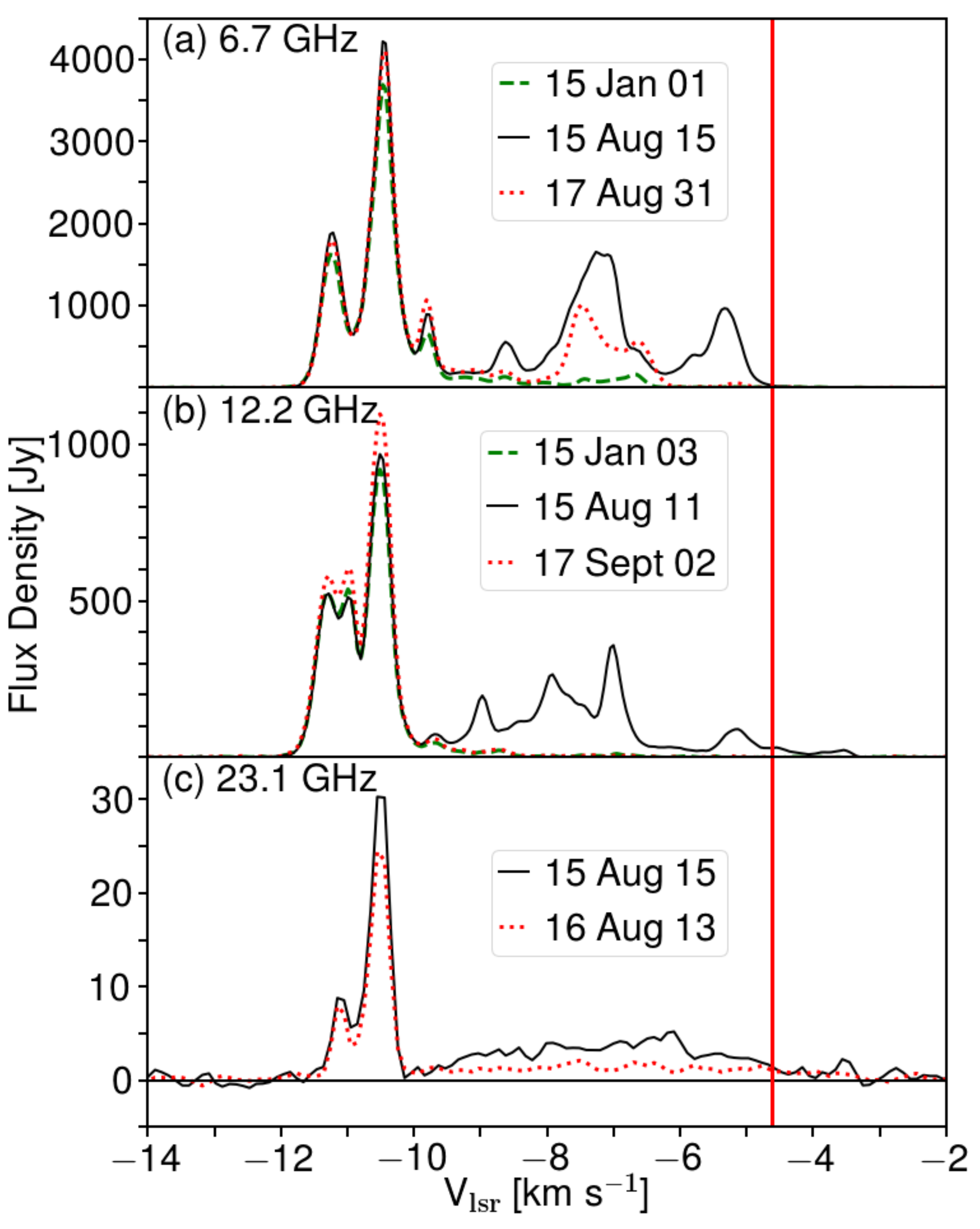}
	\caption{Spectra of methanol masers associated with \ngci\ at or near onset of Kitty (dashed green spectra), at or near the peak of Kitty (solid black spectra), and recent observations (dotted red spectra). Methanol spectra are presented for three transitions: (a) 6.7\,GHz, (b) 12.2\,GHz, and (c) 23.1\,GHz. Note that the 2016 August 13 spectra for 23.1\,GHz methanol is an average of the 13 observations taken from 2016 April 15 to 2017 January 01. The vertical red line demarcates the last red shifted channel with 6.7\,GHz emission.}
 \label{fig:meth_sp}
\end{figure}

The vertical lines in the dynamic spectra indicate times for which spectra are plotted in Fig.~\ref{fig:meth_sp}. The top (a) and middle (b) plots are for 6.7 and 12.2\,GHz respectively, while the bottom (c) plot shows spectra for two epochs from the 23.1\,GHz line of \meth. In the latter panel, the spectrum labelled 2015 August 15 is an average of spectra taken over three days around the given date. The 2016 August 13 spectrum is an average of 13 observations taken between 2016 April 15 and 2017 January 01. In all of these spectra, the profiles are complex and cannot be resolved (in velocity space) into individual features of maser emission. They are dominated by the components with $v < -10$\kms\ which display variability, but are similar to the spectra when 6.7\,GHz emission was first discovered in \ngci\ by \citet{m91} and in 12.2 and 23.1\,GHz by \citet{mb89}. Between velocities of $-10$ and $-4.6$\kms, the 6.7\,GHz emission shows considerable variation, as does the 12.2\,GHz line but that has emission with velocity components that extend further to $v \sim -3.25$\kms. There was also an increase in the 23.1\,GHz emission in this velocity range.

\begin{figure}   
	\includegraphics[width=\columnwidth]{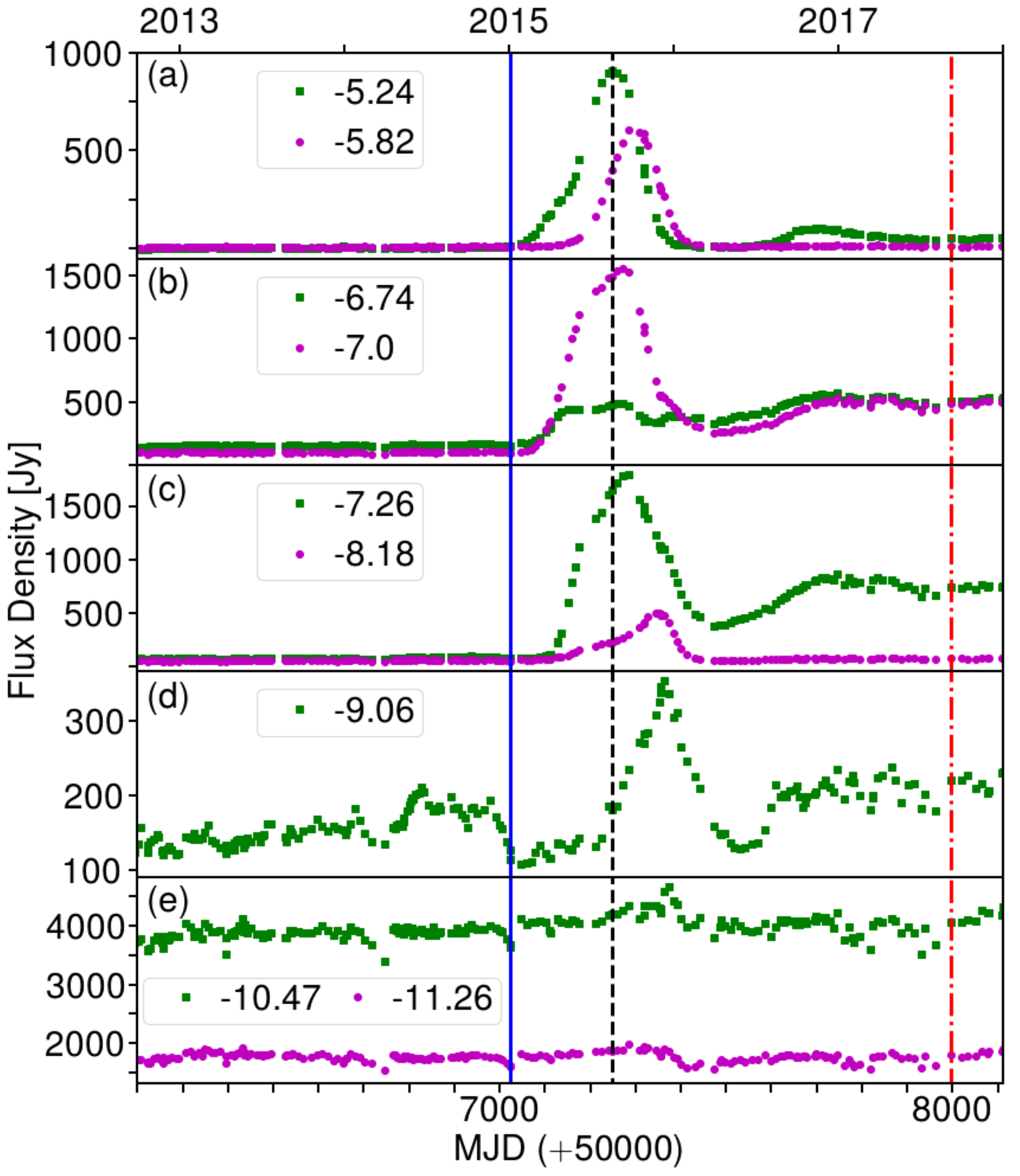}
	\caption{Time series of the flux density for selected velocity channels (in \kms) of 6.7\,GHz methanol spectra. The vertical lines are defined in Fig.~\ref{fig:m67_ct}.}
 \label{fig:m67_ts_Kitty}
\end{figure}

To investigate the temporal behaviour of the flare, we examined individual velocity channels in the spectra and generated time series plots. For the 6.7\,GHz methanol masers we present the time series plots for selected velocity channels in Fig.~\ref{fig:m67_ts_Kitty}. We show a variety of channels displaying different evolutionary patterns. The profiles in (a) are roughly symmetrical but the start and peak times occur at different times. In (b) the profiles are asymmetric and both display significant emission after minimum.  The $v = -7.26$\kms\/ channel in (c) was the component with the strongest flux density, while the other component reached its peak later and decays very rapidly.  In (d) the shape of the flare has a triangular shape which is different to the other profiles, while in (e) which has velocity components with $v < -10$\kms\/ there is very little variation associated with the flare.  This stable behaviour suggests that these velocity components are in regions outside the influence of the flare, or the flaring components are too weak to be detected.  The asymmetric shapes of the light curves shown in panels (b) and (c) of Fig.~\ref{fig:m67_ts_Kitty} can be due to more than one spot of maser emission turning on at different times in a particular velocity channel.  

The start time of the flare in each velocity channel was determined by looking for the first data point that lay immediately above the quiescent level, and thereafter monotonically increased, while peaks represent the date with the maximum measured flux density. The cadence of the observations, which varied depending on telescope availability, determines the uncertainty in the timing measurements. This method was applied to all transitions studied here. The results for each velocity channel for all the transitions that were observed, are presented in Table~\ref{tab:Kitty}.

\begin{table}
\caption{Flare information of individual maser velocity channels associated with Kitty. The time lags of flare onsets and peaks of channels are determined against the onset (2015 January 01/MJD 57023 = day 0) of the $-$5.24\kms\ 6.7\,GHz methanol maser velocity channel. Estimated error margins are shown in parenthesis.}
\label{tab:Kitty}
\begin{tabular}{ccccc}
\hline
 Vel. & \multicolumn{2}{c}{Offsets} & \multicolumn{2}{c}{Characteristics} \\
      & Onset & Peak & $F_{Peak}$  & $F_{peak}/F_{onset}$ \\
(\kms)&(days) &(days)& (Jy)        & \\
 \multicolumn{5}{c}{6.7 GHz Methanol Masers}\\
 $-$4.72 & 38(13) & 236(11) & 48 & $>$17 \\ 
 $-$5.24 & 0(1) & 226(9) & 907 & 95 \\ 
 $-$5.33 & 0(1) & 236(11) & 968 & 124 \\
 $-$5.73 & 88(1) & 263(19) & 558 & 46 \\ 
 $-$5.82 & 79(14) & 263(19) & 609 & 86 \\  
 $-$6.74 & 0(1) & 236(11) & 493 & 3 \\ 
 $-$7.00 & 0(1) & 248(14) & 1558 & 16 \\
 $-$7.26 & 47(9) & 263(19) & 1793 & 21 \\   
 $-$7.53 & 56(9) & 263(19) & 1316 & 29 \\   
 $-$7.70 & 0(1) & 263(19) & 920 & 26 \\   
 $-$7.92 & 47(9) & 263(19) & 526 & 9 \\   
 $-$8.62 & 0(1) & 263(19) & 621 & 4 \\        
 $-$9.06 & 189(36) & 340(9) & 353 & 3 \\   
 $-$9.28 & 189(36) & 295(5) & 493 & 4 \\   
 $-$9.50 & 88(1) & 370(5) & 347  & 3 \\
\hline
\multicolumn{5}{c}{12.2 GHz Methanol Masers}\\
 $-$3.54 & 142(7) & 255(4) & 30 & 13 \\
 $-$3.83 & 155(14) & 262(14) & 21 & $>$21 \\
 $-$4.55 & 110(6) & 282(16) & 64 & $>$60 \\
 $-$5.12 & 80(12) & 255(4) & 107 & $>$107 \\
 $-$5.65 & 89(9) & 255(4) & 30 & $>$16 \\
 $-$6.18 & 110(6) & 282(16) & 63 & 23 \\
 $-$7.00 & 80(12) & 255(4) & 387 & 66 \\
 $-$7.91 & 89(9) & 255(4) & 297 & 145 \\
 $-$8.64 & 80(12) & 262(14) & 106 & 11 \\
 $-$8.97 & 80(12) & 255(4) & 377 & 39 \\
 $-$9.69 & 80(12) & 262(14) & 102 & 3 \\
\hline
\multicolumn{5}{c}{22.2 GHz Water Masers}\\
 $-$3.00  & 83(21) & 337(9)  & 1420 & 34 \\ 
 $-$5.00  & 22(10) & 229(11)  & 2158 & 15 \\ 
 $-$7.11  & 22(10) & 219(13)  & 12109 & 20 \\ 
 $-$8.37  & 55(9) & 219(13)  & 5491 & 27 \\ 
 $-$24.49 & 62(14) & 344(9) & 1133 & $>$304 \\
\hline
\multicolumn{5}{c}{1665 MHz Hydroxyl Masers}\\
 $-$7.40R & 188(45) & 309(7) & 32 & 4 \\
 $-$8.10L & 113(14) & 396(8) & 205 & 26 \\
\hline
\end{tabular} \end{table}
		  
From Table~\ref{tab:Kitty} it can be seen that flaring began in several 6.7\,GHz methanol velocity channels on 2015 January 01 (MJD 57023); we refer to this date as `day 0'. Fortuitously, an observation with a null detection was recorded on the previous day so the uncertainty in the onset of the flare in our data is one day. This is the date we have used as the starting time for the flare and represented by the vertical line (blue in the online version) in Figs.~\ref{fig:m67_ct} and \ref{fig:m67_ts_Kitty}.
Based on time of onset, there are four groupings of components associated with Kitty, those that (1) began on 2015 January 01, (2) about 50\,d later, (3) about 85\,d later, and (4) about 190\,d later. For comparison, a fifth set of velocity channels not associated with Kitty are plotted. The first velocity channel of Kitty to reach a maximum was $-$5.24\kms, then there is a weak correlation between progressively blue shifted velocity channels and increasing date of peak flux density in the velocity range $-6.1$ to $-5.1$. We see a similar progression between $-7.8$ to $-6.8$\kms; no such progression is obvious in the velocity range $-10$ to $-8.0$\kms. For some of the 6.7\,GHz velocity channels the flare was roughly symmetrical; the temporal behaviour is 220$\pm$50\,d for the rise and 170$\pm$30\,d for the decay and lasted for a total of 385$\pm$50\,d. A simple measure of variability, e.g. $F_{peak}/F_{onset}$, is also listed for each component in Table~\ref{tab:Kitty}. This ratio varied from 3 to 124 for all the channels associated with Kitty. The strongest feature at $-$7.26\kms\ increased by a factor of 21. Several velocity channels continue to flare.

On average, at our sensitivity levels, the 12.2\,GHz velocity channels began flaring about 6 weeks after their 6.7\,GHz counterparts, but they all reached peak emission within a few weeks of each other. No observations were taken between 2016 January to August; this made it difficult to determine the termination dates of some channels or if they experienced secondary flares. The average emission rise times of the 12.2\,GHz channels are 160$\pm$25\,d and we estimate the decay of some are 95$\pm$20\,d. The ratio of flare peak emission to quiescent values in the 12.2\,GHz channels varied from 3 to 145 and the strongest component at $-$7.0\kms\ increased by a factor of 66, three times more than its 6.7\,GHz counterpart.

\subsection{Water Masers}
\begin{figure*}
	\includegraphics[width=\textwidth]{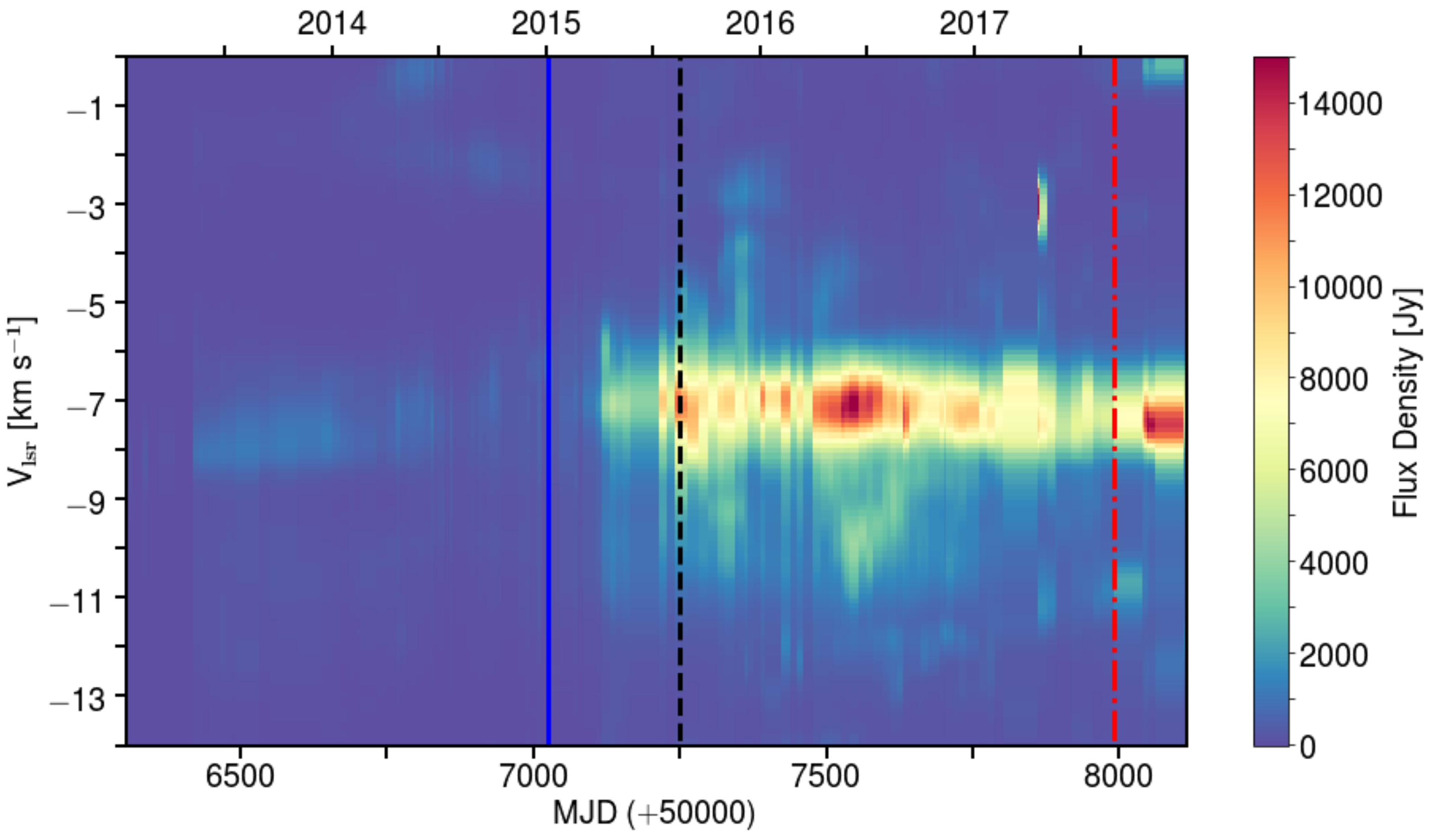} 
	\caption{Dynamic spectrum of 22.2\,GHz water masers in \ngci. The solid, dashed and dot-dashed vertical lines are defined in Fig.~\ref{fig:m67_ct}.}
 \label{fig:h2o_ct}
\end{figure*}

A dynamic spectrum made from the regular monitoring data of the 22.2\,GHz line from \water\ masers is shown in Fig.~\ref{fig:h2o_ct}. Because the flux scale in this plot reaches 15\,000\,Jy, it is difficult to distinguish features below $\sim$1\,000\,Jy. There is some emission at the level of $< 1\,200$\,Jy between $\sim -7$ and $-8$\kms\ in 2014, but this drops to $\sim 300$\,Jy prior to the start of 2015. The first indication of the flare occurred on 2015 January 23 (MJD 57046). The first peak, which reached $F$(22.2) = 12\,100\,Jy, occurred at approximately the same time as the first peak in the methanol 6.7\,GHz data (see Table~\ref{tab:Kitty}). The water masers rebrightened in mid-2016, reaching another peak of $F$(22.2) = 15\,200\,Jy. In 2017 March a short burst occurred at $v = -3$\kms\ which we have classified as a super-outburst (MacLeod et al. in preparation 2018). At the end of 2017 another brightening phase occurred, during which the peak flux density reached 15\,800\,Jy.

\begin{figure}
	\includegraphics[width=\columnwidth]{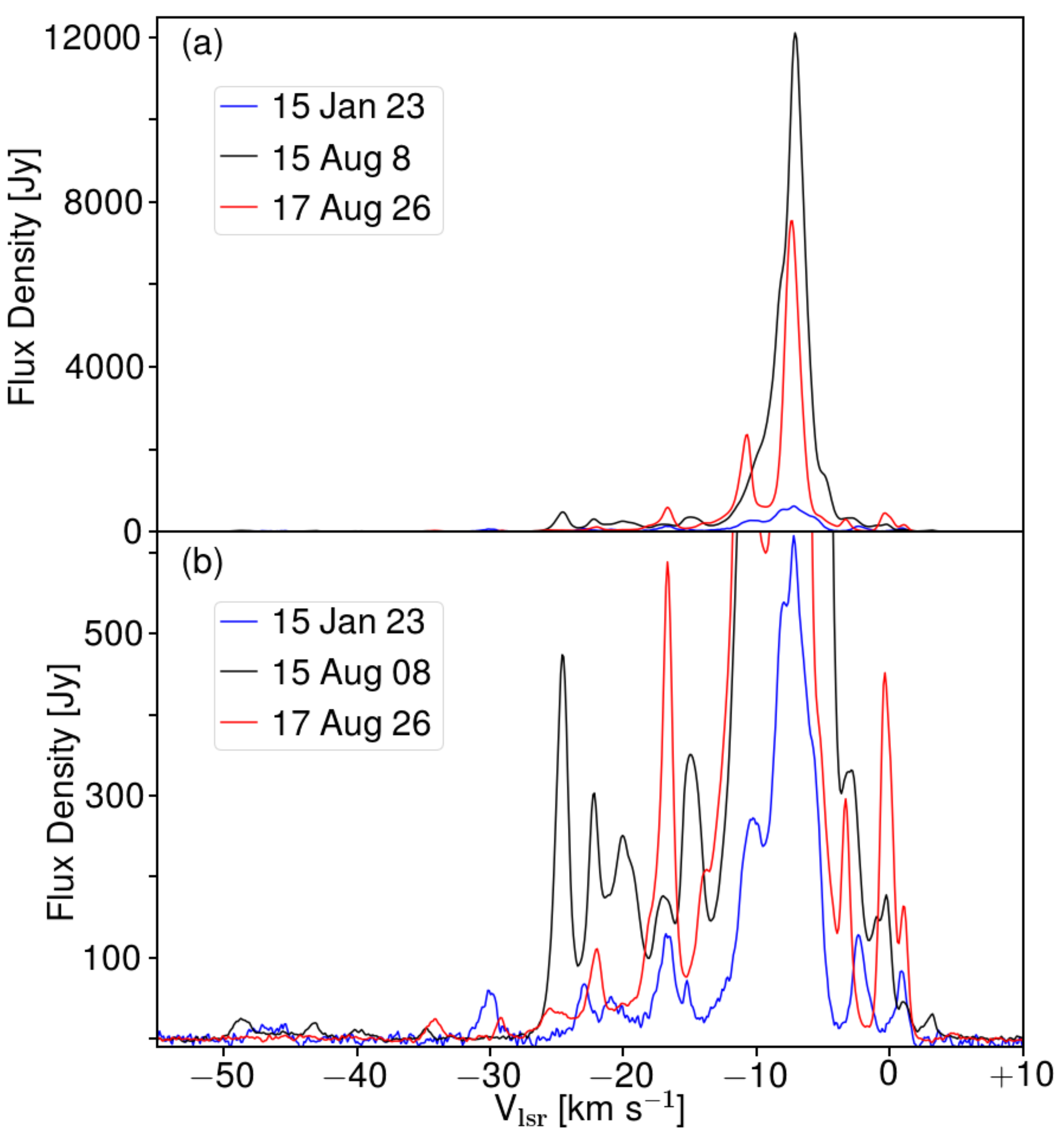} 
	\caption{Spectra of water masers associated with \ngci\ at the onset (2015 January 23) of Kitty (blue), at or near the peak (2015 August 08) of Kitty (black), and recent (2017 August 26) observations (red). The entire flux density range is shown in (a) and components with flux density less than 625\,Jy in (b).}
\label{fig:h2o_sp}
\end{figure}

The emission from water masers in \ngci\ has a larger velocity range than the methanol masers, as can be seen in the spectra shown in Fig.~\ref{fig:h2o_sp}. Maser emission occurs from $v = -50$ to above 0\kms\ in this region with all of the velocity channels showing variability on some time scale. The spectra are complex, consisting of the emission of numerous sources in the beam that cannot be resolved into individual velocity components. The peak flux density in 2015 occurred in the $v = -7.11$\kms\ velocity channel, which is similar to the 6.7 and 12.2\,GHz methanol lines that peaked at $v=-7.26$ and $-7.00$\kms, respectively. The flux in this channel increased by a factor of 20 which is similar to the 6.7\,GHz \meth\ line. The water velocity channel $v = -24.49$\kms\ varied the most (of all transitions observed), it increased by a factor of $>$300 and is located in CM2 (Brogan et al. 2018, in preparation).

\subsection{Hydroxyl Masers}
\begin{figure*}
	\includegraphics[clip, width=\textwidth]{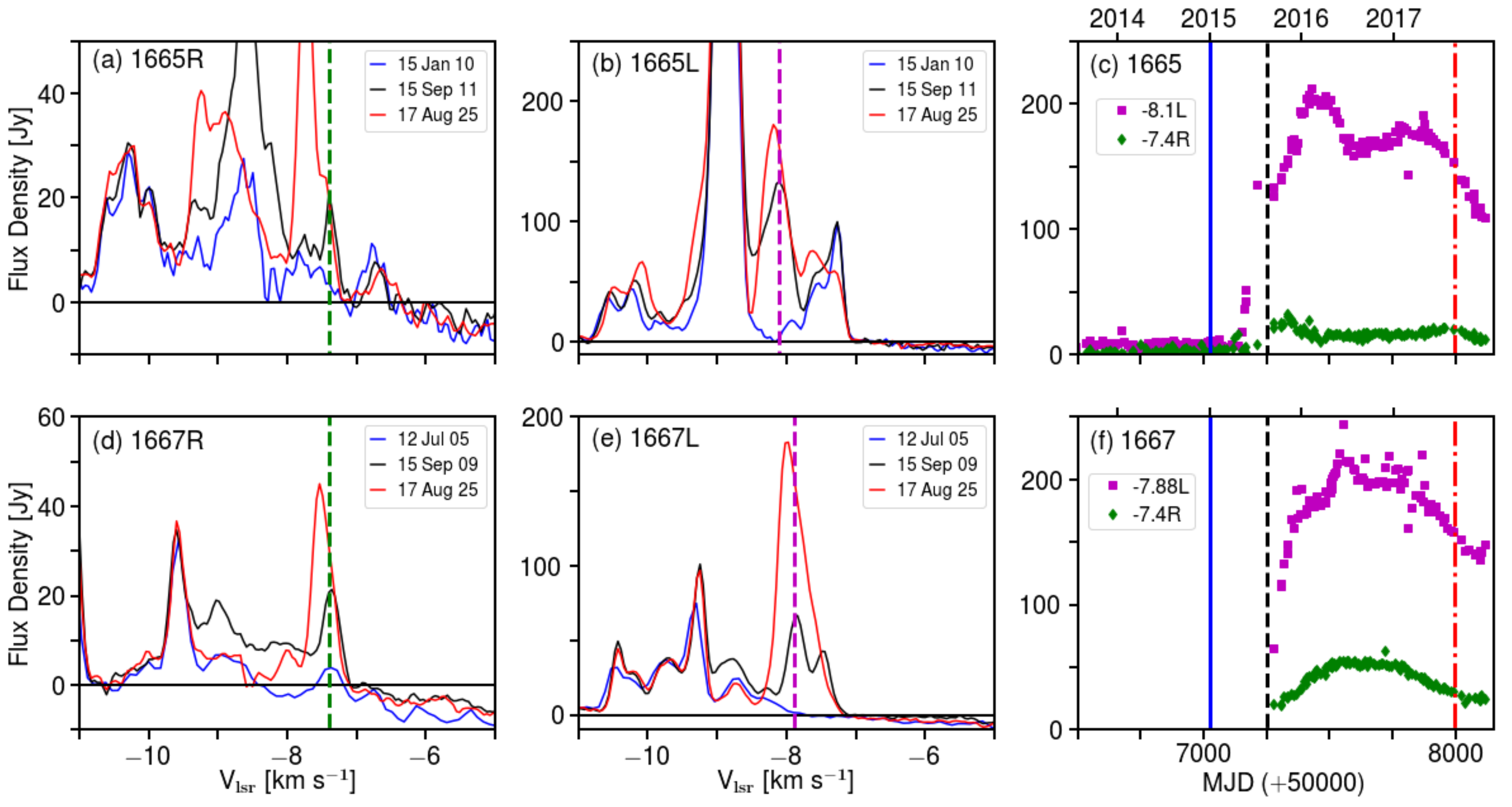} 
	\caption{Spectra of the 1665 and 1667\,MHz OH masers at or near onset, at or near peak, and recent observations of Kitty. Also time series of selected velocity channels are plotted. (a) RCP spectra, (b) LCP spectra, (c) RCP and LCP time series are presented for 1665\,MHz and similarly for 1667\,MHz in (d), (e) and (f). The vertical lines in (c) and (f) are defined in Fig.~\ref{fig:m67_ct} while the velocity channels plotted are those identified by the dashed vertical lines in the RCP and LCP spectra plots. In each spectrum an absorption feature, the result of a foreground cloud, is present at $v > -7$ \kms.}
 \label{fig:oh1665_ts}
\end{figure*}

The spectra of the 1665\,MHz groundstate OH masers are complex and contain many features that are variable and have been present since before 2013. Absorption features, at $v > -7$\kms, from foreground clouds are seen in each spectrum. The LCP spectrum had a velocity spread of $-7$ to $-11$\kms\ and the brightest flux density measured prior to 2015 was $\sim 500$\,Jy, whereas the RCP spectrum was spread over $v = -6$ to $-13$\kms and peaked at $\sim 100$\,Jy. To identify components associated with this flaring event, dynamic spectra were made (not shown) and we looked for any velocity channels with sudden increases in brightness around the time of the methanol flare. Only one channel in each polarisation was definitively identified. In LCP flaring activity started in the $v = -8.1$\kms\ channel 113 days after 2015 January 01, and in RCP the $v=-7.40$\kms\ channel shows some activity after 188 days. Spectra nearest to the commencement, peak and a recent time are shown in Fig.~\ref{fig:oh1665_ts}, together with time series plots of the identified channels.  

 The regular monitoring programme did not include  1667\,MHz groundstate OH observations and, hence, we only have data after 2015 September (MJD 57275), which is 251 days after the start of Kitty. As for the 1665\,MHz lines, the 1667\,MHz LCP and RCP spectra are complex, but both occur over velocities from $-15$ to $-7$\kms. There are some consistent features (based on spectra obtained on 2012 July 05/MJD 56114), and some variable components. RCP and LCP spectra are shown in Fig.~\ref{fig:oh1665_ts} (d) and (e) respectively, and time series (f) of the two velocity channels that varied around the time of the flare and attained the highest peak in each polarisation. The flux densities in the spectra, like those for 1665\,MHz, are stronger in LCP than RCP\@. Note that neighbouring velocity channels also varied, but their peaks were smaller than those plotted, although after the event in some cases they have become stronger than the chosen channels. The peaks in the 1667\,MHz spectra occur even later than those in the 1665\,MHz spectra.
 
 \begin{figure*}
	\includegraphics[width=0.9\textwidth]{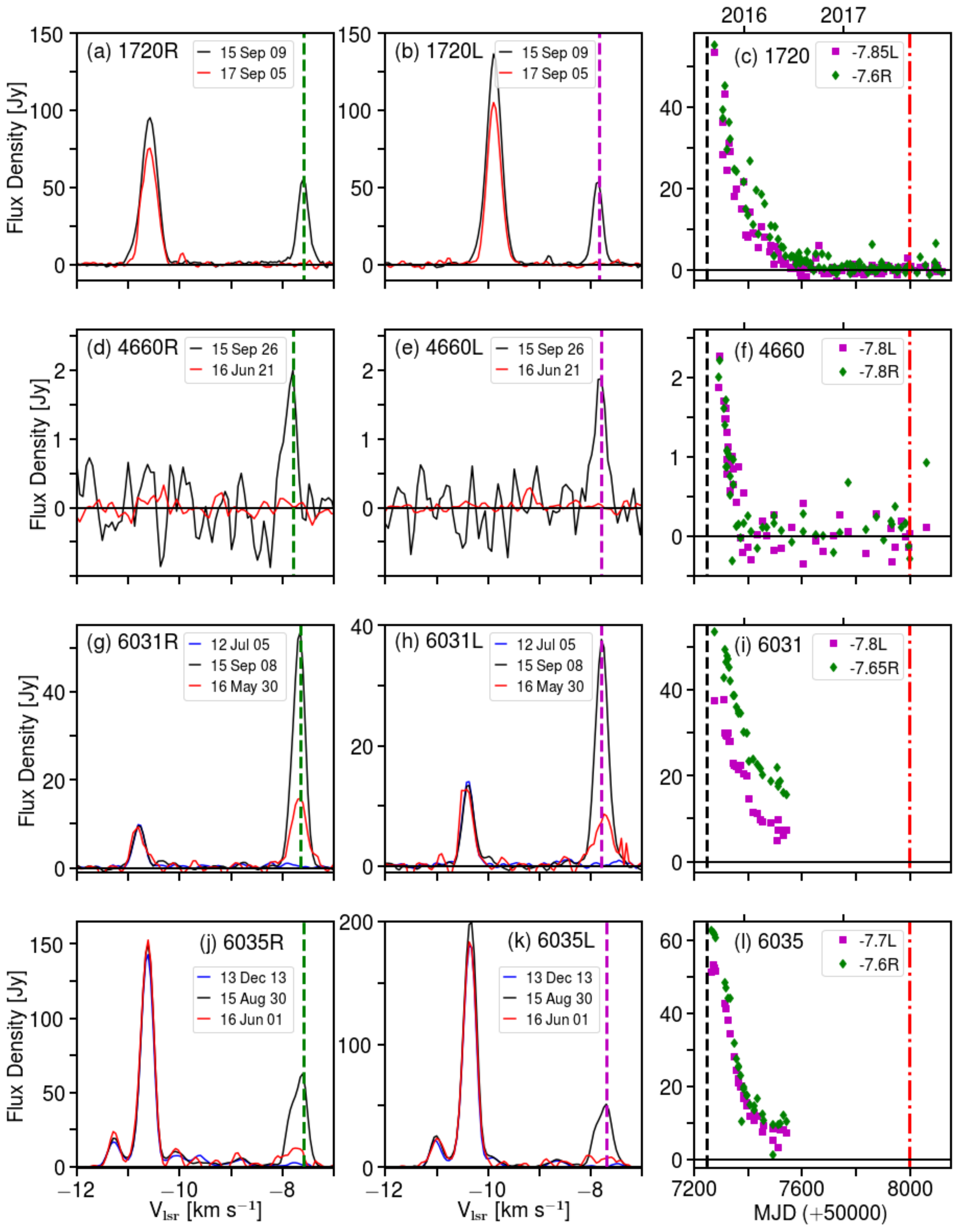}
	\caption{Spectra of the various OH transitions' observations taken: near or before 2015 January 01 (if available), nearest to 2015 August 15, and recent are presented for 1720\,MHz in (a) RCP and (b) LCP. Time series plots of selected velocity channels for 1720\,MHz masers are shown in (c). Similar plots for 4660\,MHz are presented in (d), (e), and (f), for 6031\,MHz in (g), (h) and (i), and for 6035\,MHz in (j), (k), and (l). The vertical line in each spectrum corresponds to the velocity channel plotted in the time series in the rightmost column. The vertical lines in the time series plots are defined in Fig.~\ref{fig:m67_ct}.}
 \label{fig:oh_other_ts}
\end{figure*} 
 
 We looked for the satellite groundstate OH lines at 1612 and 1720\,MHz. Nothing was detected at 1612\,MHz over multiple epochs. A 1720\,MHz observation on 2015 September 09 revealed a previously known Zeeman pair at $v = -10.2$\kms\ and a new feature at $v = -7.7$\kms\ in RCP and LCP with similar flux densities. In Fig.~\ref{fig:oh_other_ts} we show spectra for RCP (a) and LCP (b) and time series (c) of the peak velocity channel in each polarisation. The time series decays monotonically, but we only started observing at, or after, the methanol peak (vertical dashed line in (c)) had been reached.  The signal disappeared below our detection limit after 247\,d. The line profile is simple and can be fitted using a single Gaussian. The velocity difference between the peak of the RCP and LCP profiles corresponds to a Zeeman splitting due to a magnetic field of $+1.3\pm 0.2$\,mG\/ using a Zeeman splitting coefficient published in \citet{fram03}.  

A surprising discovery was the detection of a 4660\,MHz excited OH maser with a flux density of 4.4\,Jy at $v =-7.8$\kms, that died away after 71\,d. The LCP and RCP observations, shown in Fig.~\ref{fig:oh_other_ts} (e) and (f), indicate no circular polarisation for this source. The time series for the peak velocity channels are shown in (g). This is only the fifth 4660\,MHz maser ever detected. Searches did not detect any 4765\,MHz emission which is usually the strongest 4.7\,GHz line found, and had been found at a velocity of $v = -10.3$\kms\ by \citet{cmc95}. The rarity of 4660\,MHz masers suggests unusual conditions for their existence. Multiple epoch observations subsequently have not detected any of the 4.7\,GHz OH masers in this source.

Masers from the first rotationally excited level of OH at 6031 and 6035\,MHz have been found in this source since their original detection by \citet{grg70}. The strongest features lie around $v = -10.5$\kms, but weaker emission has been reported in the range from $v = -12$ to $-7$\kms\ \citep{cv95}. In the RCP and LCP 6031 and 6035\,MHz spectra shown in Figs.~\ref{fig:oh_other_ts} (g), (h), (j), and (k) respectively, these persistent features can be seen. Significant new peaks had developed around $v \sim -7.7$\kms\ when we observed them on 2015 September 08 (MJD 57274) and August 30 (MJD 57265). The time series of the peak velocity channels at $v = -7.8$, $-7.65$, $-7.7$ and $-7.6$\kms\ are shown in (i) and (l) respectively, and show that the flux density decayed over the following several months. Unfortunately, due to radio frequency interference (RFI), observations became untenable after 2016 June, which explains the truncated time series. 

The peak flux density of the new feature in the 6031\,MHz spectra was higher initially than the persistent lines around $v = -10.5$\kms. The flux density of the 6035\,MHz lines were slightly stronger than the 6031\,MHz lines but were much weaker than the long-term lines. In both of these excited OH lines the LCP flux density was consistently weaker than the RCP, the same as for the 1720\,MHz lines, whereas for the OH mainlines (see Fig.~\ref{fig:oh1665_ts}) the LCP flux density was significantly stronger than RCP\@. The asymmetric shape of the 6035\,MHz profiles at $v = -7.7$\kms\ indicates that there is more than one component of maser emission. 

\subsection{An earlier event}
\begin{figure}
	\includegraphics[width=\columnwidth]{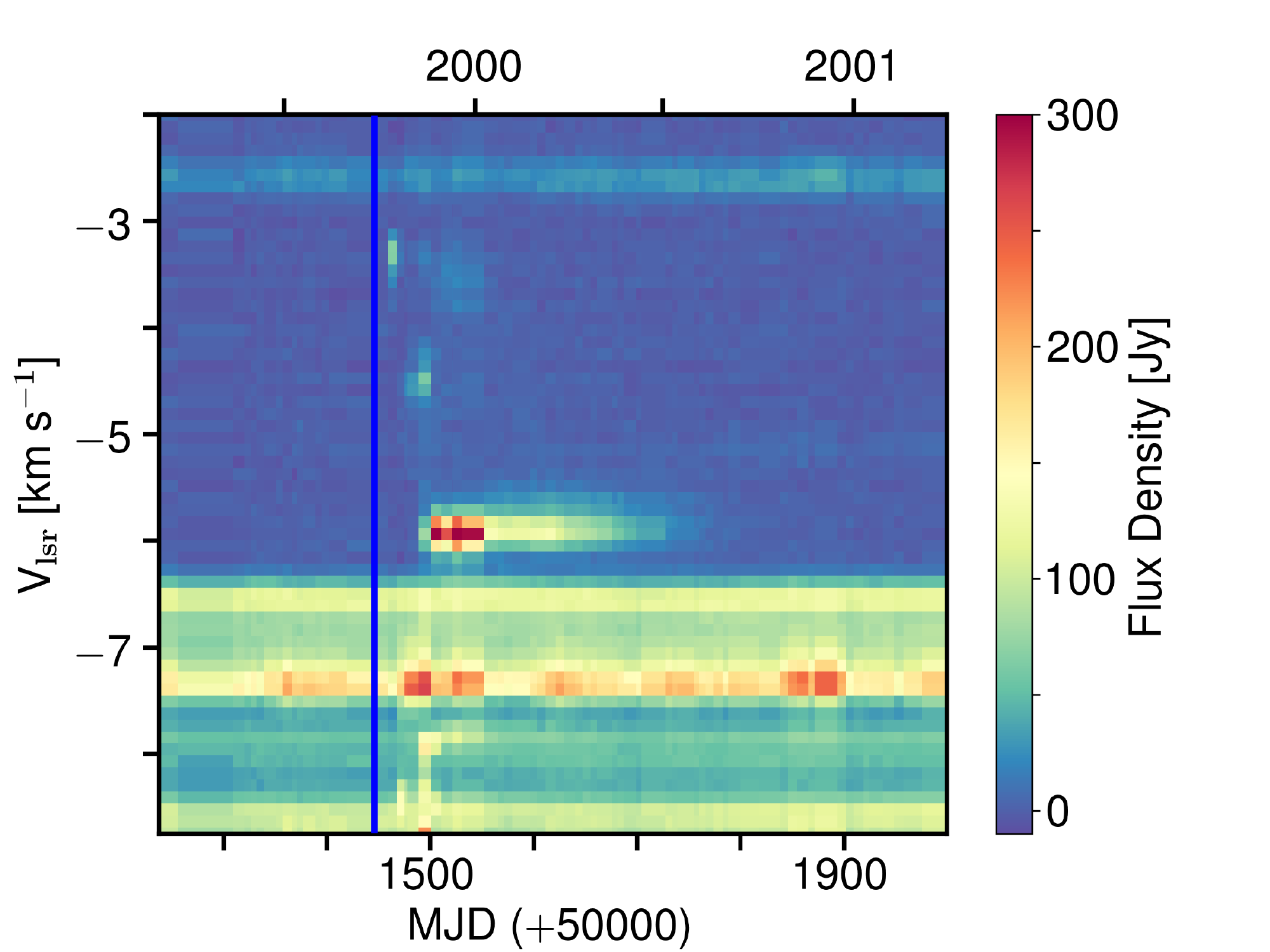}
	\caption{Dynamic spectrum of 6.7\,GHz methanol masers in \ngci. The solid vertical line is the start time, 1999 September 25 (MJD 51447), of a small flaring event (Mini) in 1999.}
 \label{fig:m67_ct_Mini}
\end{figure}

We have re-analysed the 6.7\,GHz methanol data for \ngci\ presented in \citet{ggv04}. A dynamic spectrum of the data from 1999 February 28 (MJD 51238) to 2001 March 31 (MJD 52000) is shown in Fig.~\ref{fig:m67_ct_Mini}. Based on an analysis of the time series for individual velocity channels (see Fig.~\ref{fig:m67_ts_Mini} in Appendix~\ref{Appendix A}), the brightest part of the flare is in velocity channel $v=-5.99$\kms, but it starts in the $v=-8.46$\kms\ channel at the time indicated by the vertical line. This event only reached a maximum flux density of $\sim 400$\,Jy and lasted for 340\,d. Channels that flared covered a velocity range from $-3$ to $-8.6$\kms, which is similar to the range for Kitty. There is no evidence of flaring in the persistent masers with $v < -8.6$\kms. Unfortunately, there is no data at other maser frequencies for this event. We refer to this event as `Mini'.

\subsection{Summary of results}
\begin{figure*}
    \centering
    \includegraphics[width=\textwidth]{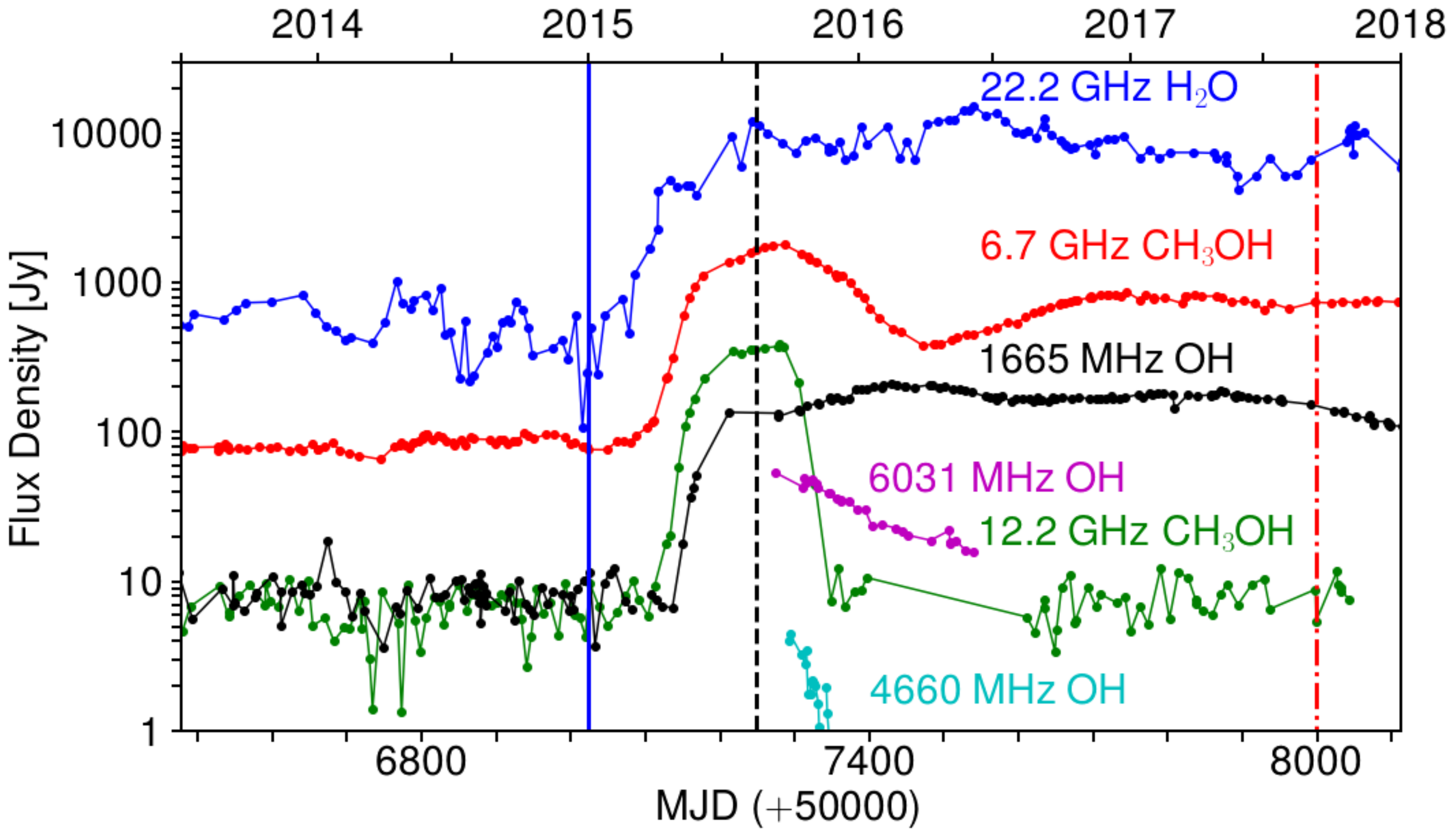}
    \caption{Time series of selected velocity channels from 6.7\,GHz ($-$7.26\kms) and 12.2\,GHz ($-$7.00\kms) methanol, 22.2\,GHz ($-$7.11\kms) water, and 1665\,MHz ($-$8.10\kms\ LCP) hydroxyl. For comparison purposes, 4660\,MHz ($-$7.80\kms\ LCP+RCP) and 6031\,MHz ($-$7.65\kms\ RCP) OH velocity channels are plotted. The vertical lines are defined in Fig.~\ref{fig:m67_ct}. Note that a logarithmic scale has been used for the flux density.}
    \label{fig:4trans_ts}
\end{figure*}

To summarise the results discussed above, in Fig.~\ref{fig:4trans_ts} we compare time series of the velocity channels for water, methanol and hydroxyl that contained the peak maser emission.  The vertical solid line indicates the start of the event as defined by one 6.7\,GHz methanol velocity channel (see Fig.~\ref{fig:m67_ct}). Data from the long term monitoring programme shows that there was some quiescent maser flux density in each of the chosen channels prior to 2015, but then the flux density increases by more than an order of magnitude in each maser line. The increases do not all occur simultaneously. The 12.2\,GHz methanol line returned to its quiescent level after $\sim 260$\,d.  The 6.7\,GHz line decayed significantly in some channels which passed through a dip before rebrightening, while the water and 1665\,MHz hydroxyl masers were active throughout the observation period reported here. The total extent of the velocity range in which all these masers resided in all transitions is $-3$ to $-10$\kms, with the exception of the high velocity water maser at $v = -24.49$\kms.

The lines of OH that were not part of a long term monitoring programme (1720/4660/6035/6031\,MHz), were only observed after 2015 August 15 (MJD 57249). They all declined monotonically after detection; the time series for selected channels of the 4660 and 6031\,MHz transitions are shown in Fig.~\ref{fig:4trans_ts} for comparison purposes.

\section{Discussion}
\label{discussion}
As a well known source of maser emission in the Galaxy, \ngci\/ has been found to host numerous masers from methanol, water and hydroxyl. The spectral appearance of some of these masers have been fairly consistent since their discovery, varying to some extent over the intervening decades. In contrast, for many other regions in the literature, the general variability of masers in all species has been reported extensively and regular monitoring programmes \citep[e.g., in water by][]{Felli07} have led to the discovery of flaring events in certain transitions. What is new about this study is that we have long term monitoring data of 6.7 and 12.2\,GHz \meth, 22.2\,GHz \water\ and 1665\,MHz OH maser lines that show contemporaneous flaring activity after a long period of relative stability. The velocities of these flaring masers all occur in the range from $v = -3$ to $-10$\kms. We also found new OH masers at 1720, 4660, 6031 and 6035\,MHz and for \meth\ at 23.1\,GHz. The 1667\,MHz groundstate OH maser follows the behaviour of the 1665\,MHz mainline, but we do not have enough data to conclusively claim a starting date for its activity.

\begin{figure*}
	\includegraphics[width=0.9\linewidth]{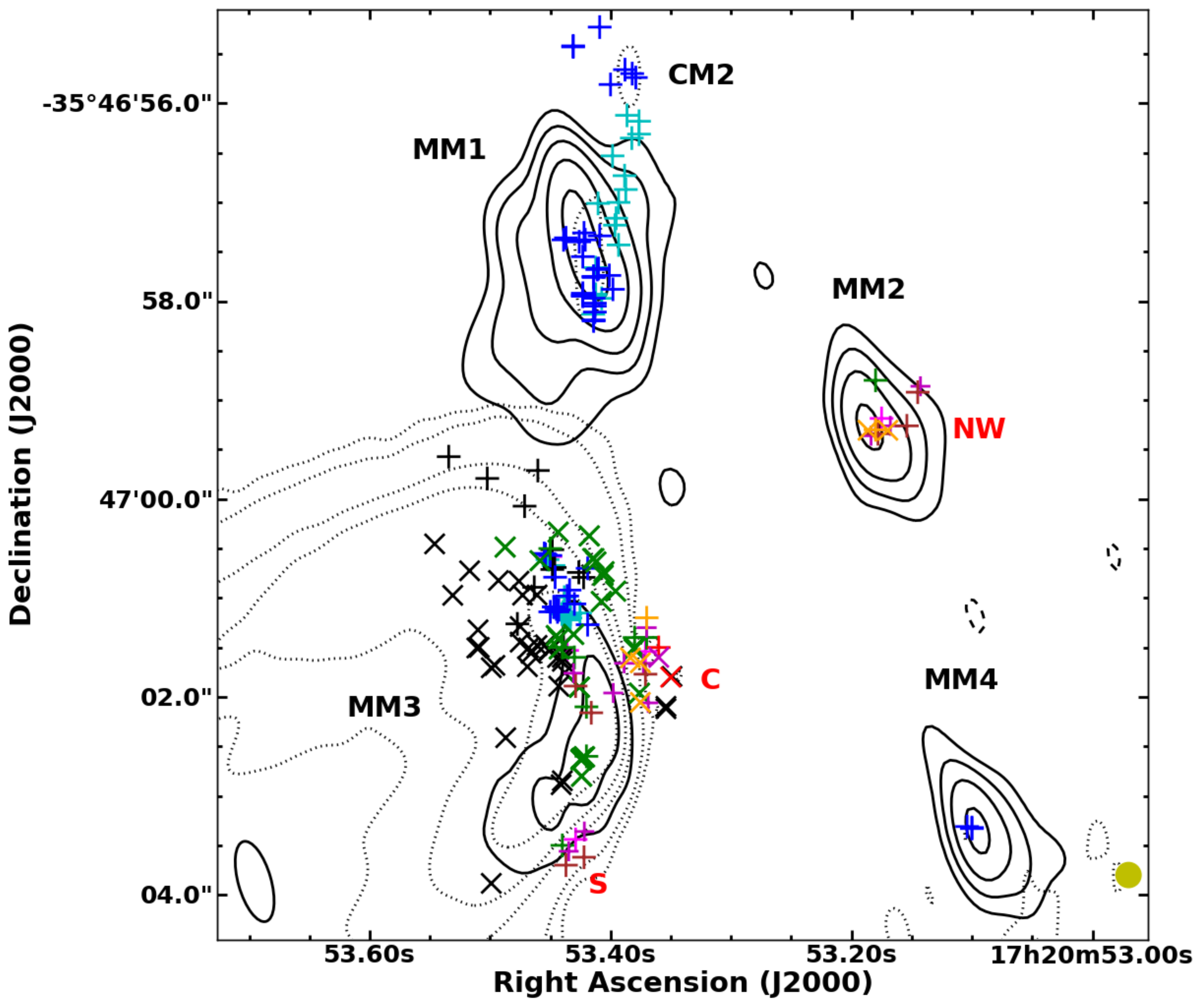} 
	\caption{Image of \ngci\ with all associated maser spots reported prior to the 2015 outburst. The SMA 1.3\,mm continuum image (epoch 2008.6) is plotted as solid contours with intensity values: $-$0.024, 0.024, 0.06, 0.12, 0.24, and 0.48\,Jy\,beam$^{-1}$ from \citet{hetal17}. The 5\,cm VLA continuum image (epoch 2011.5) is plotted as dotted contours with intensity values: 0.012, 0.6, 6.0, 12.0, 24.0\,mJy~beam$^{-1}$ from \citet{bhccfi16}. Maser spots presented are: 6.7\,GHz methanol masers from \citet{kevb13} and \citet{bhccfi16} (magenta x and + respectively), 6.7\,GHz methanol masers from \citet{Walsh98} (brown +), 12.2\,GHz methanol masers from \citet{becgvfqa12} (orange +), 19.9 and 23.1\,GHz methanol masers from \citet{kevb13} (red + and x respectively), 44\,GHz methanol masers from \citet{glhkha10} (yellow dot), 22.2\,GHz water masers from \citet{fc89} and \citet{bhccfi16} (cyan and blue + respectively), OH masers from \citet{fc89} and \citet{arm00} (black + and x respectively), and excited OH masers from \citet{gcm15} (green +) and \citet{ckr11} (black x). The \citet{fc89} water positions have been shifted by $-0\farcs7,-1\farcs0$ for the reason described in \citet{bhccfi16}. The 5\,cm source (CM2) and the four millimetre source (MM1$-$4) are labeled in black; MM3 is the \UCHII\ region \ngcf. \citet{elletal96} identified 3 regions of methanol maser activity, C, S and NW, labeled in red.  The new 6.7\,GHz methanol masers that appeared toward and surrounding MM1 and CM2 after the outburst (not shown) are reported by \citet{hunter18}.
	}
 \label{fig:spots}
\end{figure*}

To orient the subsequent discussion, a comprehensive list of maser positions from various interferometric measurements made prior to the outburst are plotted in Fig.~\ref{fig:spots}, overlaid with 1.3\,mm dust continuum emission observed in 2008 by \citet{hetal17} with the Submillimeter Array (SMA), and 6\,cm continuum observed in 2011 with the National Science Foundation's Karl G. Jansky Very Large Array (VLA) by \citet{bhccfi16}.  Using ALMA, 
\citet{bhccfi16} found that of the four primary millimetre sources that make up \ngci, the brightest, MM1, is comprised of several compact components named MM1A, B, etc. These observations were taken coincidentally when the methanol masers initially peaked in 2015 August. \citet{hetal17} reported that the dust luminosity of MM1 increased by a factor of $\sim$70 between 2011 May and 2015 July/August with the increase centered on MM1B. We report the maser emission similarly increased by factors of 10s to 100s between 2015 January and August (see Table~\ref{tab:Kitty}) suggesting a direct relationship. In recent VLA observations, \citet{hunter18} confirmed our suspicion that these flaring masers arise from new spatial regions of maser activity. In particular, they report six new regions of 6.7\,GHz methanol maser activity with overlapping velocity ranges, four in MM1 and two in CM2, and new 6.0\,GHz excited OH maser activity associated with CM2.

For the purpose of our discussion below we take the distance to \ngci\ to be 1.3\,kpc based on maser parallax measurements of the adjacent core NGC~6334I(N) \citep{chibueze14,Reid14}.

\subsection{Temporal behaviour}
Six velocity channels in the 6.7\,GHz \meth\ spectra started flaring on 2015 January 01 (day 0 in Table~\ref{tab:Kitty} and indicated by the solid vertical line in Figs.~\ref{fig:m67_ct}, \ref{fig:m67_ts_Kitty}, \ref{fig:h2o_ct}, \ref{fig:oh1665_ts}, and \ref{fig:4trans_ts}). The other 6.7\,GHz channels all started flaring later than this.  About half of the channels reached their peak on day 263, some of which started flaring after day 0. The earliest date on which the 12.2\,GHz methanol masers started flaring was on day 80, and more than half of the velocity channels peaked on day 255. Given the cadence of the observations, it appears that the 6.7 and 12.2\,GHz \meth\ masers peaked around the same time. Observations of the excited OH masers and the 1720\,MHz groundstate OH masers only started around day 250, but then decayed monotonically. These observations are consistent with these OH masers peaking around the same time as the methanol masers. The 22.2\,GHz \water\ masers started flaring on day 22 and have remained strong. Because of the rapid fluctuations in these masers, there is no clear date when they reached a peak. The OH 1665\,MHz LCP and RCP channels only started flaring 113 and 188\,d after day 0. The 1665 and 1667\,MHz groundstate OH masers have remained strong with no obvious peak in the light curve.

After peaking some of the 6.7\,GHz methanol decayed back to quiescent levels, while others went through a rebrightening. The 12.2\,GHz methanol velocity channels decayed rapidly back to their quiescent levels with no rebrightening. If the methanol masers are radiatively pumped by the object in MM1B then the 12.2\,GHz masers only start flaring when the source becomes bright or hot while the 6.7\,GHz masers are pumped over a broader range of conditions. The rebrightening of the 6.7\,GHz methanol masers could be due to another (weaker) flaring event in MM1B, which does not pump the 12.2\,GHz masers. The 6.7\,GHz flare in 1999 was possibly also due to a similar type of event which was weaker than this flare. The start of flaring in some velocity channels after day 0 can be explained by conditions in the cloud not supporting masers pointing towards us, and, likewise, the different times at which the masers peaked could be due to geometrical effects or the masers dying out before the source peaked. 

The methanol spot maps presented in \citet{hunter18} were taken in 2016 November (MJD 57711) during the peak of the rebrightening of the methanol masers reported here. The dominant new maser in their observations, at $v = -7.25$\kms, is about 1000\,au in projected distance from MM1B. Other areas of MM1 are at projected distances of between 650 to 2800\,au from MM1B. If MM1B is the source of this flaring event, as proposed by \citet{hetal17}, then the projected light travel time ranges from 4 to 16\,d; much shorter than our estimated time lags. Similarly, the light travel time to MM2 (3$\farcs$5) and MM3 ($4\farcs$5) is 25 and 33\,d in the sky plane, but there is no evidence of associated flaring in the masers of the latter two regions, even after 1100\,d.  It is possible that the geometric dissipation of the energy in the event was sufficient that it had no impact on the masers in MM2 and MM3.  Somewhat less likely is that MM2 and MM3 have not yet witnessed the event, but this requires them to be at least 0.5\,pc in the background relative to MM1. Regardless, the new masers in MM1 and CM2 are highly variable in comparison to the decades long stable masers in MM2 and MM3, possibly the result of the energy injected into it by the event at MM1B causing instabilities in the masing conditions.

\subsection{6.0\,GHz excited OH masers}
Maps of the 6.0\,GHz excited OH masers in \ngci\ made by \citet{ckr11} and \citet{hunter18} show that the persistent masers around $v \sim -10.5$\kms\ are located in MM3. These masers have not shown any variations associated with the 2015 flare. 

The new masers we detected at $v \sim -7.7$\kms\ in 2015 September were not present in single dish observations in 2012 or 2013. We do not have data for their onset, but observed them from at (or near) their peak, after which they decayed. There is a gap of $\sim$160\,d between the termination of our single dish observations and the VLA observations of \citet{hunter18}, who found new 6.0\,GHz masers at the same velocity ($-7.7$\kms) in CM2 that are stronger than when the single dish observations ceased. It is not clear whether the flares in our data are from CM2, which had rebrightened by the time of the VLA observations, or from some region closer to MM1 that have subsequently faded below the sensitivity limit of the VLA. If these masers are radiatively pumped, then the light curves we obtained and the CM2 masers in the VLA observations could be the same. \citet{hunter18} detected a new 6.7\,GHz methanol feature in CM2 at $v = -7.6$\kms, a similar velocity to the 6.0\,GHz OH masers, increasing our confidence of their relationship to Kitty. We note that a temporal correlation between 6.7\,GHz methanol and 6.0\,GHz OH masers in a flaring event has been reported in the past in a massive protostar \citep[IRAS~18566+0408,][]{AlMarzouk12}.

\subsection{4.7\,GHz OH masers}
A 4765\,MHz excited OH maser feature with a flux density of 0.3\,Jy at $-$10.3\kms\ was detected in 1991 by \citet{cmc95} but it was not detected in 2000 by \citet{de02} down to a level of 0.09\,Jy. Unfortunately, the position determined has very large error bars which cover more than the whole map shown in Fig.~\ref{fig:spots}. Neither group detected emission at 4660 or 4750\,MHz. We did not detect 4750 or 4765\,MHz emission. 

The 4660\,MHz line, which is not subject to observable Zeeman shifts in weak magnetic fields, has a velocity $v = -7.8$\kms, which is similar to the 1720, 6031 and 6035\,MHz masers (which can be Zeeman shifted by a few 0.1\kms). Although we did not see these masers turn on, we assume that all these OH lines are formed in the same region and are part of the 2015 flaring event. The VLA observations of \citet{hunter18} included a high resolution spectral window on the 4660\,MHz line, but did not detect it.

What is unusual about this maser is that 4660\,MHz masers are rare (only four are reported in the literature) while $\sim30$ masers in the 4765\,MHz line have been found \citep{Qiao14}. In general, models predict that flux densities $F(4765) > F(4750) > F(4660)$. Clearly, this is not the case in Kitty. Also there has been some discussion in the literature suggesting correlations between 4765\,MHz excited OH masers and either the ground state 1720 or 1612\,MHz lines, depending on the density and temperature of the gas. Our observations show no evidence of 4765\,MHz emission relative to 1720\,MHz masers by a factor of $\sim 200$.

The 4660\,MHz emission died out rapidly, faster than the 6\,GHz excited OH and 1720\,MHz lines (see Figs.~\ref{fig:4trans_ts} and \ref{fig:oh_trans_ts}). This behavior possibly suggests that the conditions under which the 4660\,MHz masers form are much more sensitive to the conditions in the gas than the other lines.

\subsection{6.7 vs. 12.2\,GHz methanol masers}
Pumping models predict that $F(6.7) > F(12.2)$ but we find examples of velocity channels in which the converse is true.  This situation occurs for velocities $v > -4.6$\kms. There are some narrow regions with very specific conditions where $F(12.2)$ can be greater than $F(6.7)$ \citep{csg05}, but in dynamic regions where the conditions are changing we would not expect these conditions to exist for any significant length of time. No interferometric observations were taken at 12.2\,GHz during the period when the 12.2\,GHz masers were detectable. \citet{hunter18} do report 6.7\,GHz masers in this velocity range but at an epoch after the 12.2\,GHz masers had faded back to their quiescent levels. The survey of 400 6.7\,GHz maser targets by \citet{becgvfqa12} found only about three percent of objects where the 12.2\,GHz methanol masers were stronger than their 6.7\,GHz counterparts.

\subsection{6.7\,GHz methanol vs 22.2\,GHz water masers}
The contemporaneous increase in the methanol and water maser flux densities provides a strong constraint on the physical explanation of their respective flares. Recently imaged with the VLA in late 2016, the flaring 6.7~GHz methanol emission is localized to the area toward and surrounding MM1 and CM2 where strong masers now appear for the first time \citep{hunter18}. The infrared photons required to pump these masers \citep{scg97a}, as well as the OH mainline masers \citep{Gray07}, naturally would have become more abundant when the surrounding dust was rapidly heated by the increase in radiation from the proposed \citep{hetal17} protostellar outburst source MM1B. In contrast, prior to the outburst, water masers were already known to be associated with both MM1 and CM2 (see Fig.~\ref{fig:spots}), a non-thermal radio continuum source located $\sim$2\,$''$ north, which had the strongest water masers but no associated compact dust emission \citep{bhccfi16}.  The projected extent of the water masers has a mechanical crossing time of many decades, for even a high velocity protostellar jet.  Therefore, it seems unlikely that the creation of new shocks containing new water masers close to the central protostar would produce a rate of increase in the water maser emission that so closely parallels that of the radiatively-pumped methanol masers (see Fig.~\ref{fig:4trans_ts}).  Instead, we find it more likely that some property of the existing water masing gas was changed by the excess radiation produced during Kitty.  One possibility is a small increase in the ionisation fraction, which can result in more intense water masers \citep{kn87}. Alternatively, the phenomenon of superradiance that has been invoked to model the methanol and water maser flares in G107.298+5.639 \citep{rh17} could be contributing to the observed behavior of the light curves.

\subsection{Possible associated earlier events}
Using single-dish observations beginning in 1965 July, \citet{wdw68} noted changes in the OH mainline intensity of \ngc\/ by almost an order of magnitude on time scales of a few days to months. Adjacent velocity channels showed great variability. In particular, their 1665\,MHz observations at $v =-8.1$ and $-7.7$\kms\ decreased significantly in amplitude between two sets of observations, after which these channels were relatively constant at low amplitude. There was no activity in the corresponding 1667\,MHz channels. The velocity range in which the 1965 flare occurred is $-$12.7 to $-$6.6\kms\ with the largest variation occurring in the $-$7.7\kms\ channel. These channels are the same as variable features reported for Kitty presented in this paper. It is possible that shortly after the discovery of OH masers in 1963 a flaring event occurred, peaking around 1965 July, on par with Kitty reported here, and ending 1965 October seen in \citet{wdw68}. 

Another flaring event that occurred in 1999 was found in the 6.7\,GHz data from the HartRAO monitoring programme. See Appendix~\ref{Appendix A} for analysis. It was a weaker and shorter-lived flare, but the velocity range and factors of increase from quiescent flux densities ($\sim$20 to 200) were similar to Kitty. Unfortunately no other transitions were observed during this event. It is possible that this event and the earlier event in \citet{wdw68} are both related to Kitty, i.e. arising from the same physical location.

\subsection{Physical Cause of the Outburst}
It is not obvious from these observations what caused the outburst that powered Kitty. Based on (sub)millimeter observations, \citet{hetal17} speculate that the unprecedented increase in dust temperature resulted in an increase in the radiative pumping of the various maser emissions in the region. While the association between the dust emission flare and the maser flares might explain their correlated timing, it does not directly explain the cause. \citet{hetal17} suggest the occurrence of an accretion event analogous to an FU Ori outburst -- a phenomenon in low-mass protostars in which they experience rapid increases in accretion luminosity followed by long decays that occur over tens of years \citep{Hartmann96}.  The first such event in a massive protostar was recently reported \citep[S255~NIRS3,][]{Caratti17} and it also coincided with an extended methanol maser flare \citep{Moscadelli17}.  As noted by \citet{hetal17}, the timescale for such an accretion outburst to heat the dust in MM1 by the observed amount is $\sim$200\,d based on the equations of \citet{jhhb13}.  Qualitatively, one should expect the different maser transitions and velocity components to turn on at different times during the course of this heating due to differences in the local geometry and opacity of the dense medium.   While it remains uncertain how many days before the first maser flare (2015 January 01) the accretion outburst began, this 
theoretical heating timescale combined with the observed 190 d spread in maser flare start dates, suggests that it was likely only up to a few weeks.  
Indeed, the VLA observations of \citet{hunter18} confirmed that 6.7\,GHz methanol and 6.0\,GHz hydroxyl masers are spatially associated with MM1 and CM2. Features of the 1665 and 1667~MHz OH and water maser spectra associated with Kitty are also consistent with the variation originating from MM1 and CM2.  

Further support for an FU Ori-like origin hypothesis for Kitty comes from a comparison with the 1720\,MHz OH maser outburst associated with the 5.5\,mag optical flaring event in the FU~Ori object V1057~Cyg that began in 1969 \citep{Herbig77}.  The OH maser emission originally detected by \citet{Lo73} declined exponentially over a two year interval and became undetectable by 1975 \citep{Andersson79}.  A subsequent OH maser outburst was observed in 1979 which then diminished by a factor of 2 over a two month period \citep{wetal81}. Several features of the methanol masers of Kitty exhibited a rapid flare followed by an equally rapid decline, an initial flare in which the rise and fall times are similar followed by a secondary flaring event, or a rise which remained high. If Mini and Kitty indeed arose from the same spatial location, then we have recorded two such events in 15\,yr, analogous to the repeated maser outbursts in V1057~Cyg.  This conclusion is further supported by the fact that 1720\,MHz OH maser flares have been imaged toward massive protostars in the past \citep[e.g. W75N,][]{Fish11}.

As an alternative origin hypothesis, a supernova can produce sufficient energy to heat the surrounding dust that then pumps the masers. It can also produce ejecta that might result in the non-thermal radio source CM2 $\sim$2\,$''$ north of MM1; however, CM2 already existed in the VLA observations of 2011.  Also, while shock-excited 1720\,MHz OH masers are often seen towards supernova remnants \citep[e.g.,][]{brogan2013,hoffman2005}, the other groundstate lines are usually found only in absorption \citep[see e.g.,][and references therein]{Green02}, unlike in Kitty.  But perhaps the most obvious characteristic that rules out a deeply-embedded supernova is the lack of new strong centimeter continuum emission.  Using previous VLA observations of the optically obscured supernova SN 2008iz in M82 as a model \citep{Brunthaler10}, the 5~GHz flux density should have reached $\sim$100~kJy after a year with a diameter of $\sim8''$, which it quite obviously did not.

Finally, \citet{hetal17} noted the possibility that a near encounter with another star or a merger might have triggered the accretion event leading up to Kitty. \citet{bz05} propose that stellar mergers in our Galaxy will occur at the same rate as the birth rate of massive stars and will produce high luminosity infrared flares. Such flares could pump hydroxyl and methanol masers and associated shocks could pump the water masers. \citet{Bally02} presents a cartoon of the possible light curve of a binary stellar merger; he predicts progressively stronger flares at each periastron passage over decades to centuries until the actual merger occurs. However, the flare profile is different than that of the time series of the flaring masers presented here; he suggests that each flare will experience a rapid rise followed by a slower decay perhaps lasting years. Mini and Kitty may represent flares resulting from successive periastron passages with the re-brightened masers being the result of other mechanisms not considered in \citet{bz05}. 

\subsection{Prediction of a future flare}
\begin{figure}
\includegraphics[width=\columnwidth]{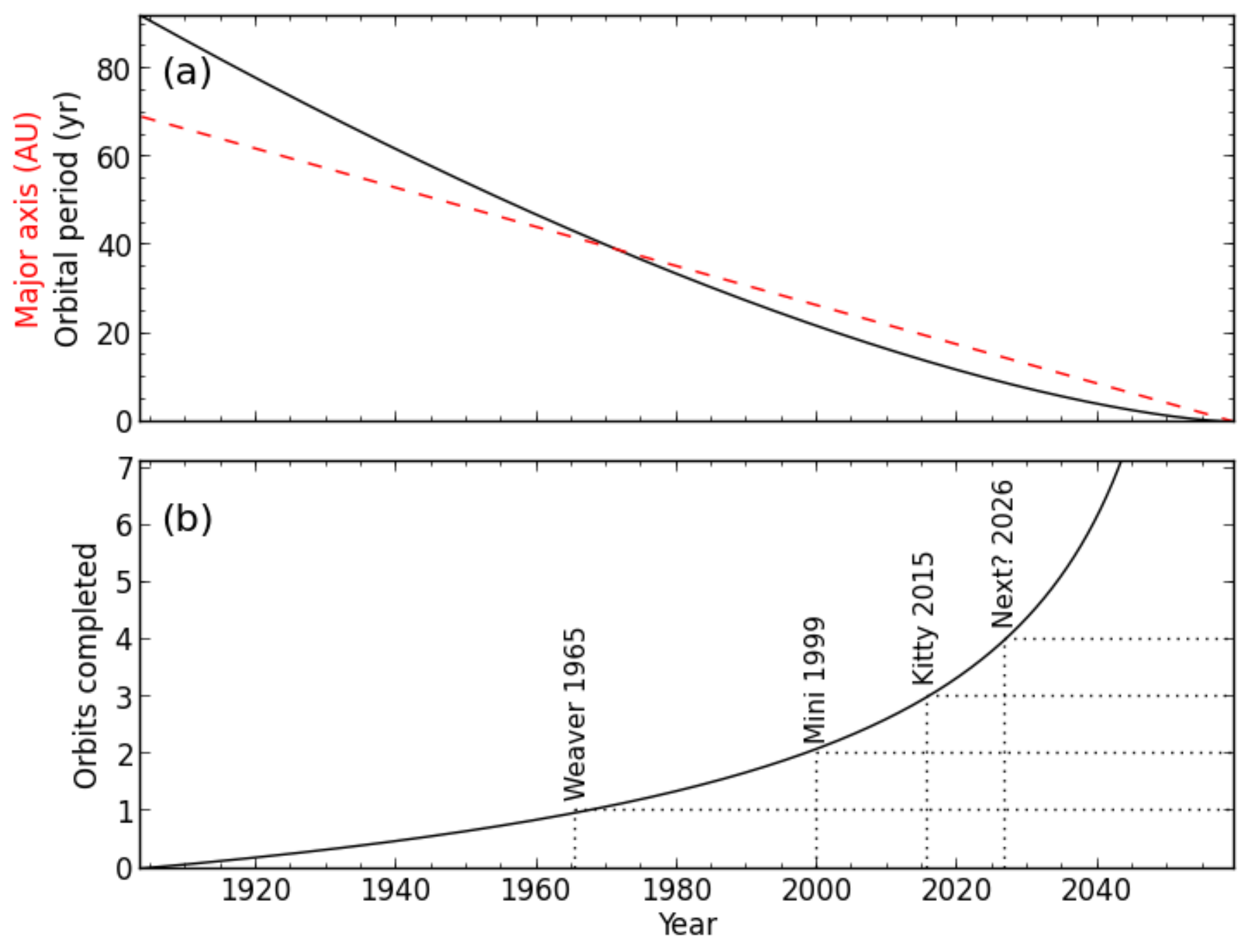}
\caption{Panel (a) shows the theoretical orbital decay of a binary system in \ngci\ using equations in \citet{s10} fit to the dates of the three known maser flares: Weaver (1965), Mini (1999), and Kitty (2015). The solid line is the orbital period in years and the dashed line is the major axis of the orbit in au. In this scenario, panel (b) shows the predicted date of the next flare to be at the next completed orbit in late 2026.}
\label{fig:orbit}
\end{figure}

If we assume the peaks of the OH flare reported by \citet{wdw68}, Mini, and Kitty occur at successive periastrons of an embedded binary star, then using equation 55 of \citet{s10} we can fit for the orbital parameters and predict the next flare event. For our best-fit solution, shown in Fig.~\ref{fig:orbit}, we estimate the coalescence time is $\sim$156\,yr, the initial orbital period is $\sim$92\,yr, and the coalescence date is in $\sim$2059.  In this scenario, the next flare is predicted to occur in late 2026. If we further assume that the cloud sound speed is $\sim$1.6\kms, and the ambient cloud H$_{2}$ density is $\sim6\times10^{7}$\,\pcmm, then using equation 56 of \citet{s10} the total mass of the system is $\sim$4.9\,M$_{\odot}$. This value is roughly consistent with a (proto) B3 zero age main sequence (ZAMS) star being the dominant member \citep{Hanson97}, and a ZAMS star of this type could power the measured properties of the hypercompact HII region surrounding MM1B \citep{bhccfi16,brogan17}.  However, the derived mass from the orbital model is strongly dependent on the assumed sound speed in the gas. In any case, \citet{s10} suggest that the energy of interactions at periastron will rival that of the combined stellar output only during the final few per cent of the merger process.  Nevertheless, we expect the next flare may be more powerful than Kitty. 

\section{Summary and Future Work}
We report here a significant flaring event in 10 transitions in three molecular species associated with \ngci\ that began 2015 January 01. The 6.7\,GHz methanol and  22.2\,GHz water masers began flaring contemporaneously within $\pm$22\,d of each other, while the 12.2\,GHz methanol and 1665\,MHz hydroxyl masers flared 80 and 113\,d later respectively. The flaring for all transitions occurred in the velocity range $-$10 to $-2$\kms. The strongest flaring methanol and water features increased, by $\sim$20 times above quiescent levels; the strongest water maser feature reached $\sim$15,000\,Jy. The weak emission in some velocity channels increased by factors up to 145. This flare coincides in time with an unprecedented increase in the millimeter continuum and inferred dust temperature reported by \citet{hetal17}. 

We report the detection of only the fifth 4660\,MHz excited OH maser. We also report new maser emission at 1667, 1720, 6031 and 6035\,MHz hydroxyl, and 23.1\,GHz methanol in \ngci.

We report an earlier flare in 1999; it was only observed at 6.7\,GHz and was at most 5 times weaker than Kitty.  We also highlight a 1965 OH maser flare reported by \citet{wdw68}. We hypothesize that these three flares could be related, analogous to the repeated OH maser flares in the FU~Ori star V1057~Cyg.  Such repeated flares could be due to the orbital decay of a binary protostar, with radiative outbursts growing in strength with each successive periastron passage.  If so, we predict a future maser flare in 2026.  

We note that future observations of this event may provide constraints on maser pumping models.  Further interferometric studies will also be fruitful, and we predict that new cluster(s) of OH masers will be detected with sufficiently high resolution observations and that evidence of features associated with Kitty may be found in historical interferometric data. Finally, we will continue to monitor this interesting source and HartRAO is upgrading its spectrometer to allow for simultaneous observations at many transitions and for more sources with even shorter cadences.

\section*{Acknowledgments}
We thank Dr. Alet de Witt and Jonathan Quick for their efforts to schedule time around various other observing programmes at HartRAO. Correspondence with Professor John Bally was greatly appreciated. We thank an anonymous referee for their comments, they improved the quality of this paper.
The National Radio Astronomy Observatory is a facility of the National Science Foundation operated under agreement by the Associated Universities, Inc.  The Submillimeter Array is a joint project between the Smithsonian Astrophysical Observatory and the Academia Sinica Institute of Astronomy and Astrophysics, and is funded by the Smithsonian Institution and the Academia Sinica. Finally, we acknowledge Dr. Michael Gaylard posthumously for his leadership in maintaining these long-term monitoring programmes at HartRAO; without which this work would not have been possible.  

\bibliographystyle{mnras}
\bibliography{Monitoring_bibl_v1} 

\appendix
\section{Relation between the 1999 and 2015 events}
\label{Appendix A}
\begin{figure}
	\includegraphics[width=0.9\columnwidth]{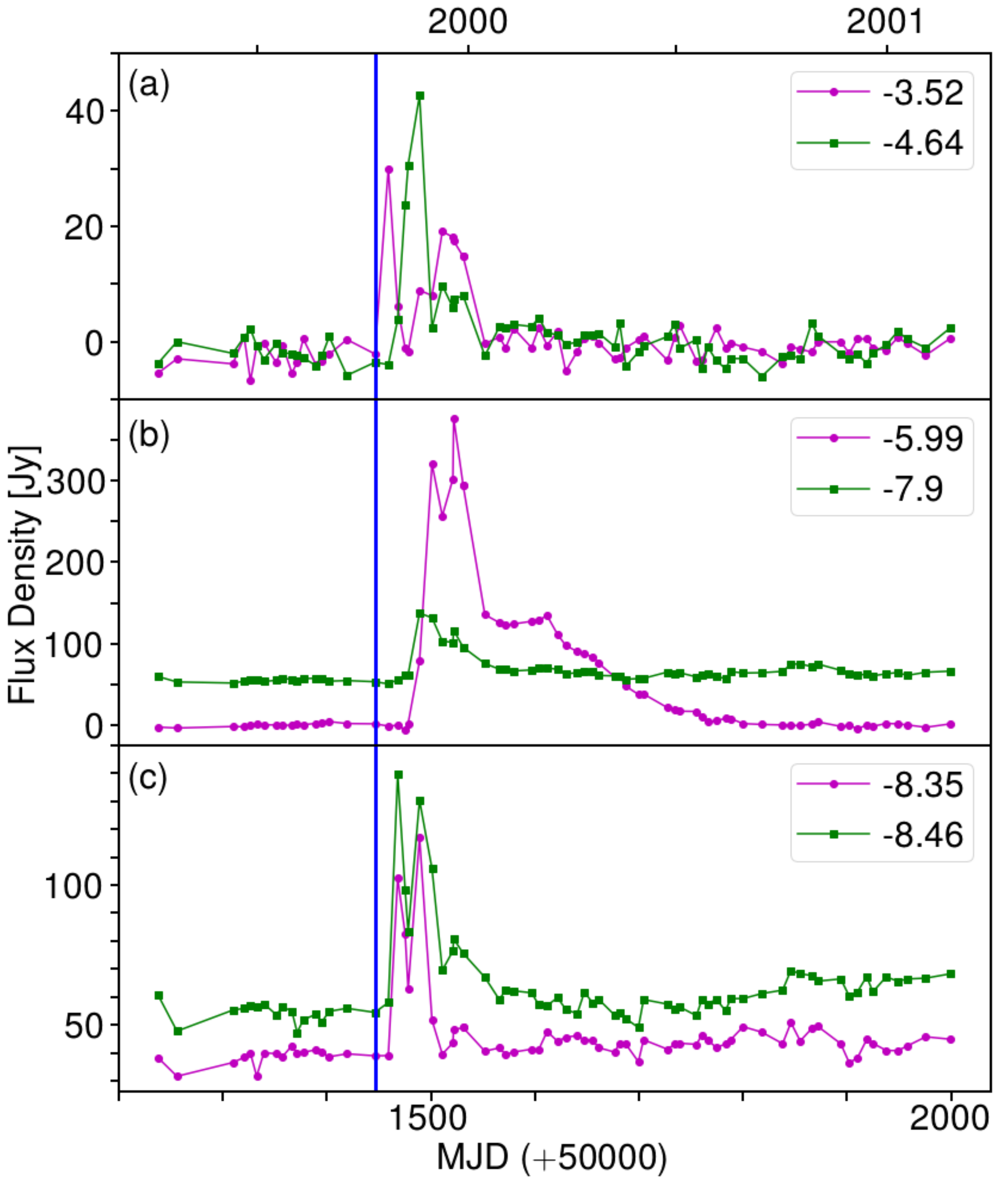}
	\caption{Time series for selected velocity channels of 6.7\,GHz methanol spectra associated with Mini. The vertical line represents the estimated onset date (1999 September 25/MJD 51447) for the $-$8.46\kms\ velocity channel.}
 \label{fig:m67_ts_Mini}
\end{figure}

Time series plots of velocity channels associated with a flaring event in 1999 (Mini) associated with \ngci\/ are shown in Fig.~\ref{fig:m67_ts_Mini} and results of our re-analysis, following the same method described for the 2015 flaring event (Kitty), are listed in Table~\ref{tab:Mini}. The onset of Mini began within $\pm$20\,d of 1999 September 25 in a number of velocity channels, while other channels turned on between 10 and 30\,d later. The strongest emission associated with Mini occurred in the $-$5.99\kms\ channel; emission in this channel turned on 32\,d after the onset at $-$8.46\kms\ and terminated 253\,d after that determined for $-$8.46\kms. Typically velocity channels associated with Mini rose quickly, between 20 and 130\,d, and declined more slowly, between 40 and 300\,d. Some velocity channels show evidence of a secondary flare, e.g. $-$3.52\kms, as occurred in channels associated with Kitty. 

\begin{figure}
	\includegraphics[width=\columnwidth]{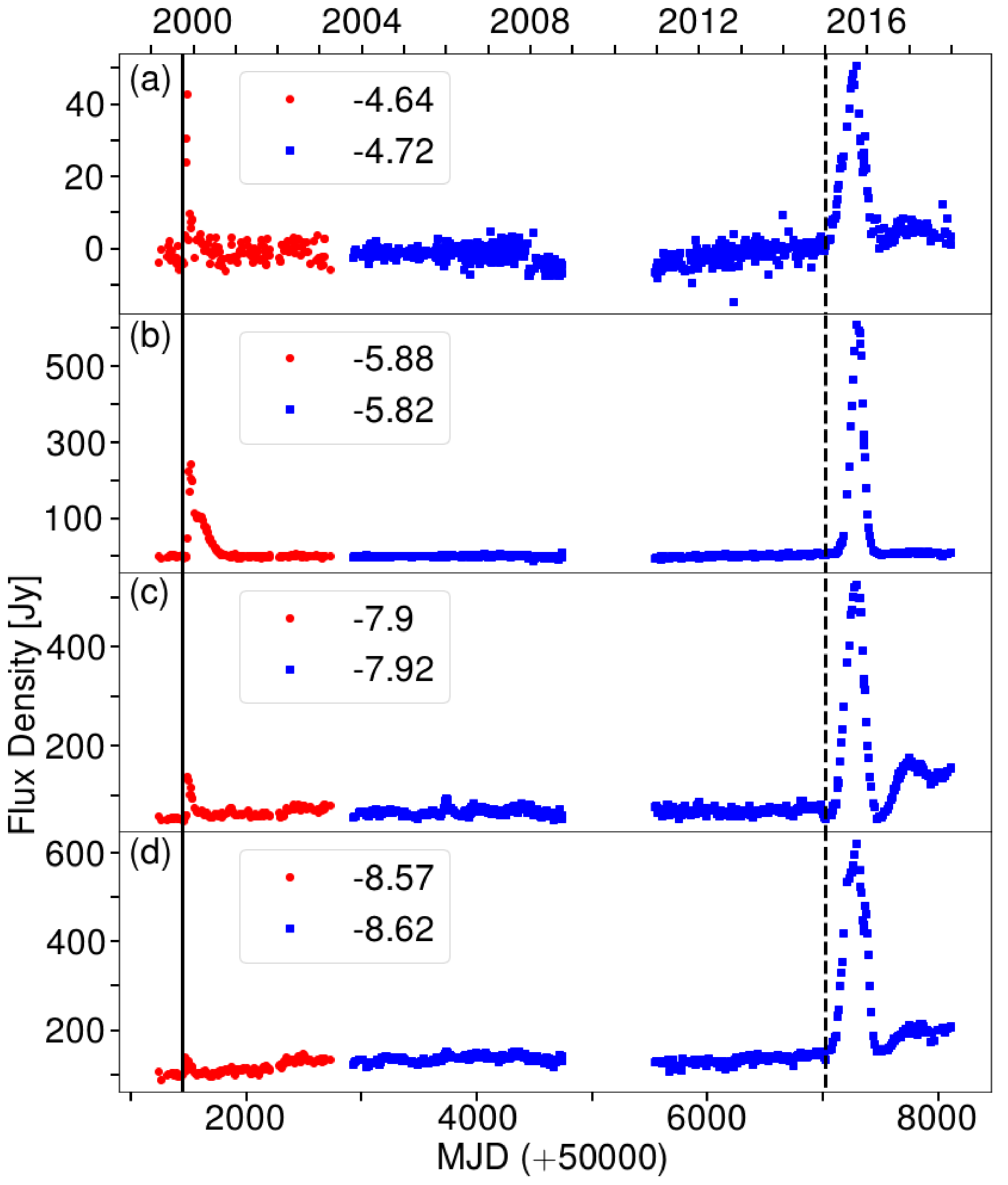}
	\caption{Time series for selected 6.7\,GHz methanol velocity channels common to both the 1999 and 2015 event; they are: (a) $-$4.7, (b) $-$5.8, (c) $-$7.9, and (d) $-$8.6\kms. The onset date of each flare is demarcated by a vertical line: solid (1999 September 25) and dashed (2015 January 01).}
 \label{fig:m67_ts_comb}
\end{figure}

We identify velocity channels with flaring activity common to both Mini and Kitty and present them in Fig.~\ref{fig:m67_ts_comb}. Kitty was relatively stronger than Mini in these common channels with a range of ratios from 1.1 to 4.4; the brightest of Kitty, $-$7.26\kms, was about 5 times brighter than its counterpart in Mini. Fewer velocity channels in Mini showed evidence of flaring activity than in Kitty. In both Mini and Kitty, from when each channel peaked, there appears a possible progression of the flare from the red to blue shifted velocity channels in the range $-$4 to $-$6\kms. In the velocity range $-$9 to $-6.5$\kms\ for Mini the converse appears to occur.

Intriguingly, for both Mini and Kitty the channels that experienced the largest increases were in the $-$6 to $-$5\,\kms\/ range. For Mini the velocity channel $-$5.99\kms\/ increased by a factor of $\sim$230 while the nearest counterpart in Kitty, $v = -5.82$\kms\/ increased by a factor of 86 (see Table~\ref{tab:Kitty}). The largest emission increase, a factor of 124, at 6.7\,GHz in the velocity channels of Kitty was at $v = -$5.33\kms. We calculated the mean positions of all emission in the velocity channels nearest to $-$5.3 and $-$6.0\kms\ in the 2016 November VLA data ($-5.30$ and $-6.05$\kms) presented in \citet{hunter18}. Both mean positions are located in MM1-Met1 \citep{hunter18} and separated by only $\sim$0$\farcs$6 or about 750\,au at 1.3\,kpc. 

\begin{table*}
\caption{Flare information of individual 6.7\,GHz methanol maser features associated with the 1999 event (Mini). The time lags of flare onsets, peaks, and terminations of velocity channels are determined against the onset of the flare on 1999 September 25 (MJD 51447) for the $-$8.46\kms\ 6.7\,GHz methanol velocity channel.}
\label{tab:Mini}
\begin{tabular}{ccccccccc}
\hline
Vel. & S$_{Peak}$ & \multicolumn{3}{c}{Flare Offsets} & \multicolumn{4}{c}{Flare Characteristics} \\
 & & Onset & Peak & Termination & Rise & Rise & Fall & Duration\\
 (\kms) & (Jy) & (days) & (days) & (days) & (days) & Factor & (days) & (days) \\
\hline
 $-$8.57 & 141 & 0 & 22 & 119 & 22  & 1.4 & 97 & 119\\
 $-$8.46 & 140 & 0 & 22 & 119 & 22  & 2.6 & 97  & 119\\   
 $-$8.35 & 117 & 13 & 43 & 105 & 30  & 3.0 & 62  & 92\\   
 $-$7.90 & 137 & 13 & 43 & 235 & 30  & 2.7 & 192  & 222\\
 $-$6.44 & 68 & 32 & 75 & 150 & 43  & 1.4 & 75  & 118\\
 $-$5.99 & 377 & 32 & 75 & 372 & 43  & 231.5 & 297  & 340\\
 $-$5.65 & 35 & 32 & 165 & 337 & 133  & 34.8 & 172  & 305\\
 $-$5.54 & 19 & 32 & 125 & 309 & 93  & 18.8 & 184  & 277\\
 $-$4.64 & 43 & 13 & 43 & 105 & 30 & 42.7 & 62 & 92\\
 $-$3.63 & 22 & 32 & 75 & 119 & 43 & 19.9 & 44 &  87\\
 $-$3.52 & 30 & 0 & 13 & 29 & 13  & 30.0 & 16  & 29\\
\hline
\end{tabular} \end{table*}

The 1999 flaring event is a weaker, shorter duration flare with fewer active velocity channels compared to the 2015 event. However, both flares occurred across a similar velocity range ($-10$ to $-$3\kms), had a similar temporal behaviour, and showed no flaring activity at velocities less than $-$10\kms. From above we know that the methanol channels that experienced the largest variations in Kitty are located in MM1-Met1; it is possible that those for Mini were also located there. We propose that these flaring events are related.

\section{Time series of hydroxyl masers}
\label{Appendix B}

\begin{figure*}
    \centering
\includegraphics[width=\textwidth]{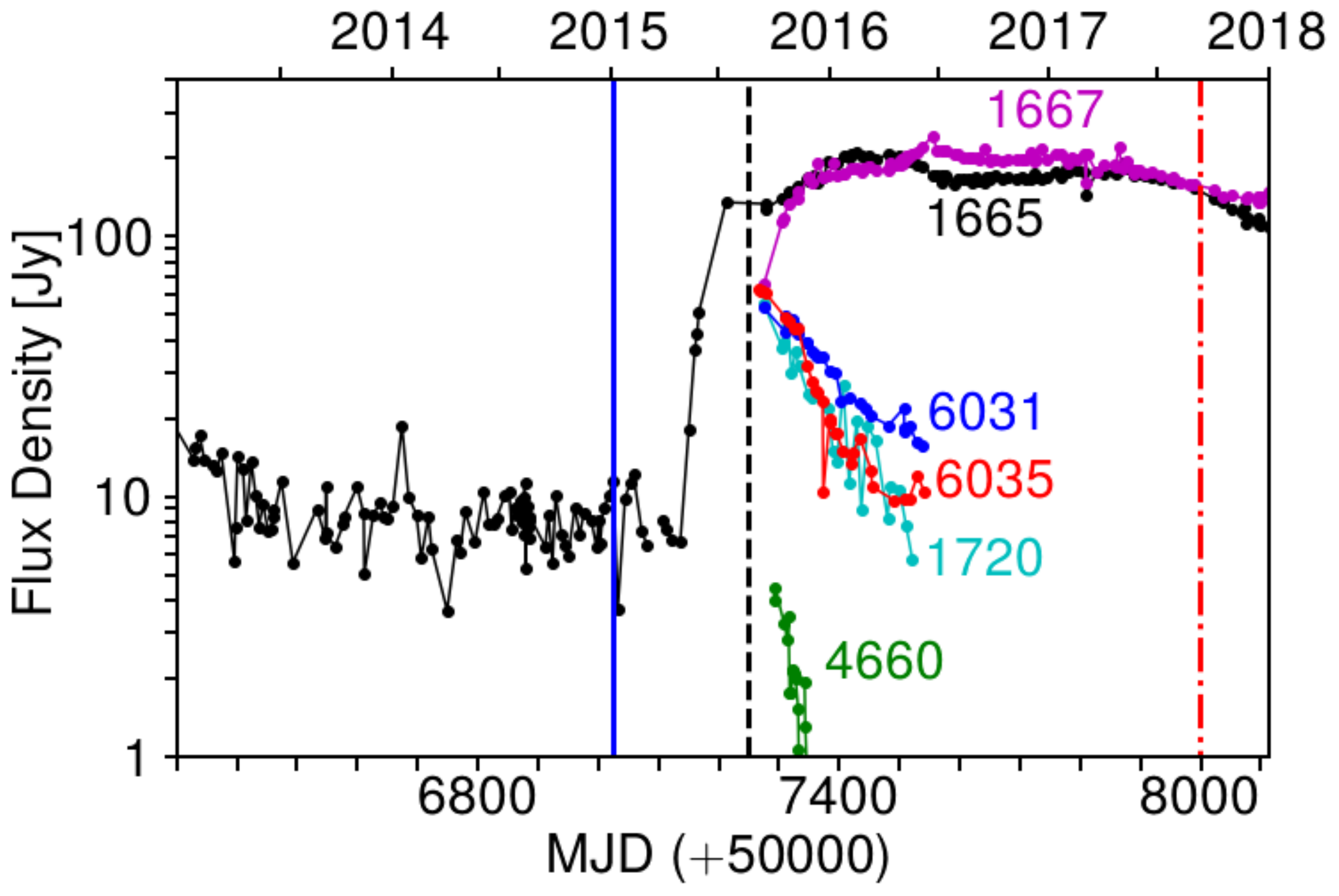}
    \caption{Time series of selected velocity channels from each of the hydroxyl transitions monitored. The OH velocity channels plotted are:
    $-$8.10\kms\ (1665\,MHz LCP), $-$7.88\kms\ (1667\,MHz LCP),
    $-$7.60\kms\ (1720\,MHz RCP),
    $-$7.80\kms\ (4660\,MHz L+RCP),
    $-$7.65\kms\ (6031\,MHz RCP), and
    $-$7.60\kms\ (6035\,MHz LCP). The vertical lines are defined in Fig.~\ref{fig:m67_ct}.
    Note that a logarithmic scale has been used for the flux density axis.}
    \label{fig:oh_trans_ts}
\end{figure*}

We present in Fig.~\ref{fig:oh_trans_ts} time series plots of selected velocity channels for each of the hydroxyl transitions observed here. We report masing at a new velocity in the 1720 and 6031\,MHz lines and much stronger maser emission in the 1667 and 6035\,MHz lines, all at $\sim-$7.7\kms. We report only the fifth 4660\,MHz maser ever detected, again at a similar velocity. 

The 1667\,MHz OH masers, like the 1665\,MHz OH masers, are contaminated by  masers unrelated to \ngci\ and by an absorption feature that is folded into the spectrum as a result of observing in frequency switching mode (at about $-16$\kms). The remaining transitions have less complex spectra and follow temporal behaviour similar to the 12.2\,GHz methanol masers with the exception of the $-$7.43\kms\ 1720\,MHz velocity channel; it peaked later. It is not clear when each of these transitions reached a maximum but we determine that they attained values of at least 10 to 135 times greater than their quiescent values. The 1720 and 4660\,MHz (see Table~\ref{tab:Kitty}) dropped below detection limits after 247 and 71\,d respectively. 

\bsp	
\label{lastpage}
\end{document}